\documentclass[useAMS,usenatbib,nodraft]{mn2e}

\usepackage[T1]{fontenc}
\usepackage{aecompl}
\usepackage{graphicx}
\usepackage{amssymb}
\usepackage{lipsum}
\usepackage{url}
\usepackage{color}

\title[Looking for activity cycles in late-type \textit{Kepler} stars]{Looking for activity cycles in late-type \textit{Kepler} stars using time--frequency analysis}
\author[K. Vida, K. Ol\'ah, \& R. Szab\'o]{
K. Vida\thanks{E-mail:vidakris@konkoly.hu}, 
K. Ol\'ah,
R. Szab\'o
\\
Research Centre for Astronomy and Earth Sciences, Hungarian Academy of Sciences}
\begin{document}

\date{Accepted 2013 XX XX. Received 2013 XX XX; in original form 2013 September XX}

\pagerange{\pageref{firstpage}--\pageref{lastpage}} \pubyear{2013}

\maketitle

\label{firstpage}

\begin{abstract}
We analyse light curves covering four years of 39 fast-rotating  ($P_\mathrm{rot}\lesssim1d$) late-type active stars from the \textit{Kepler} database. Using time--frequency analysis  (Short-Term Fourier-Transform), we find hints for activity cycles of 300--900 days at 9 targets from the changing typical latitude of the starspots, which, with the differential rotation of the stellar surface change the observed rotation period over the activity cycle. We also give a lowest estimation for the shear parameter of the differential rotation, which is $\approx 0.001$ for the cycling targets. These results populate the less studied, short period end of the rotation--cycle length relation.
\end{abstract}


\section{Introduction}

The 11 year and other, longer activity cycles of the Sun have been known for a long time, and similar multiple cycles have  been recovered for many active stars (see \citealt{2009AA...501..703O}  and references therein). Systematic studies of stellar cycles began with the   Ca H\&K survey at Mount Wilson \citep{1968ApJ...153..221W}, and with the advent of the Automated Photometric Telescopes (APTs, see e.g. \citealt{1997PASP..109..697S}). To study this phenomenon, long-term observations are needed, as the typical timescales of the cycles range from a few years to decades. A correlation has been found between the rotation period and the length of the activity cycle, as shown first by \cite{1996ApJ...460..848B} and recently by \cite{rot-cyc}, namely, that on faster rotating stars the activity cycles tend to be shorter. 

Planned, and currently developing long-term all-sky surveys, as the Large Synoptic Survey Telescope (LSST, see \citealt{lsst}) the PASS \citep{PASS}, and the Fly's Eye Camera System \citep{flyseye} may give a huge thrust to this field, provided they run for many years, even decades, since not only selected objects (which may contain no suitable targets and miss interesting ones)  but the whole sky will be monitored.

At present though, the best option for monitoring stars is the \textit{Kepler} space telescope, providing an almost continuous dataset of unprecedented precision from about 160\,000 targets, among them thousands of active stars. The already recovered stellar cycles range from years to decades, while the  \textit{Kepler} space telescope operated only for four years. \cite{shortcyc} analysed long-term photometric measurements of ultrafast-rotating ($P_\mathrm{rot}\approx 0.5d$) M-dwarfs, and found activity cycles on three stars with cycle lengths between 300--500 days, which is already within the reach of \textit{Kepler}.

 Magnetic activity,  rotation and differential rotation of \textit{Kepler} stars is in the focus of research lately (see e.g. \citealt{kepleract1,kepleract2,kepleract3}).
\cite{keplerrot} developed a robust method using an autocorrelation function to determine rotation periods from light curves and studied more than 2400 stars. 
The method was applied later by \cite{keplerrot2} to \textit{Kepler} Objects of Interest (KOIs) to study exoplanet-hosting systems.
\cite{keplerrot3} used Lomb--Scargle periodograms to search for rotation periods in \textit{Kepler} targets, the authors analyzed 12\,000 F, G and K-type stars. 
The same method was applied by \cite{keplerdr} to study more than 40\,000 active \textit{Kepler} stars, they also made an effort to estimate the values of the differential rotation shear.

Working with \textit{Kepler} data has its own drawbacks. There are instrumental trends during each observing quarter, and shorter term instrumental glitches are also present on a timescale of a few days. Although there are attempts to correct these trends (and not just remove them automatically), the possibility of having a homogeneous light curve ranging many observing quarters seems really hard to achieve. Hence, the information on long-term cycles, which can be basically seen by naked eye on a persistently observed earth-borne light curve from photometry, is lost in the \textit{Kepler} data. 

Or is it? Are there other properties of the activity cycles, that can help us to trace them?
One realisation of the 11-year cycle on the Sun is the butterfly-diagram, i.e., the phenomenon that the sunspots tend to appear on higher latitudes at the beginning of the cycle, and closer to the equator at its end.
There have been attempts to recover this migration of typical spot latitudes on other active stars (see e.g. \citealt{2007ApJ...659L.157B}). Lately, \cite{Katsova:2010kt} used wavelet analysis of Ca H\&K data from the Mount Wilson survey to determine the variations in the rotation period of active stars, and studied similar cyclic variation of the solar corona, from a ``Sun as a star'' approach.

Recent theoretical dynamo models are able to describe the  butterfly diagram reliably for different kinds of active stars.
Figure 10. in \cite{emre} shows the modelled spot distribution of a fast-rotating ($P_\mathrm{rot}=2 d$) K0V star during its activity cycles. According to the model, the well known shape of the butterfly diagram changes substantially as a result of the fast rotation. The flux tubes -- even if started from a similar configuration as the Sun -- emerging from the tachocline reach the photosphere at much higher latitudes than on the Sun. Unfortunately, in this fast rotating, late-type stellar case, the latitude distribution also changes: the spots appear in a much thinner latitude stripe (between about $35-45^\circ$) in the model,  in contrast with observed solar case (between about $0-40^\circ$). Compared to the Sun, a much smaller modulation of the emerging latitudes is still found in the model during the activity cycle, but most of the emerging spots appear almost at the same latitude. Thus the butterfly diagram in this case resembles much less to a butterfly, as the effect is almost washed away by the overlapping ``butterfly wings" (cf. Figure 9. in \citealt{emre}). 

With a strong enough differential rotation however, there might be a difference in the rotational period even in the case of the small latitudinal migration of the fast rotating active dwarf stars that is still large enough to be detected by high-precision long-term photometry with good time coverage, such as data of the \textit{Kepler} survey. The signal we have to look for is a small change in the typical latitude of spot emergence of a differentially rotating stellar surface, which would result in a very small quasi-periodic change in the photometric rotational period during an activity cycle. Long-term change of the rotational periods can be revealed by time-series period searching methods. Such effect has already been detected on the long-period giant active star CZ CVn by \cite{tifran-cycle}.

In this paper we look for cycles on fast rotating active dwarf stars through the systematic changes of their observed rotational periods due to differential rotation.

\section{Target selection}
\label{sect:targets}

\begin{table*}
\centering
\begin{tabular}{llllllllll}
KIC ID & Kepler&\multicolumn{2}{c}{Contamination (\%)}&$T_\mathrm{eff}$ & $\log g$ & $P_\mathrm{rot}$ & $P_\mathrm{cyc}$ & $\Delta P_\mathrm{rot}$ &$\alpha_\mathrm{min}$\\
&mag.&min.&max.&&&(d)&(d)&(\%)&\\
\hline

03541346 & 15.379 & 11.4 & 16.4 		    & 4194 & 4.503 & 0.9082& 330 (50)	   & .25 & .0027\\ 
04819564 & 14.672 & 3.7 & 7.8 	       	    & 4125 & 4.511 & 0.3808& 530	(150)     & .06 & .0016\\ 
04953358 & 15.487 & 14.1 & 26.5$^\dagger$& 3843 & 4.608 & 0.6490& 600 (200)/L$^\ddagger$	   & .05 & .0008\\
05791720 & 14.067 & 2.2 & 6.0 		    & 3533 & 4.132 & 0.7651& 320	(60)     & .12 & .0016\\
06675318 & 15.242 & 3.1 & 9.3 		    & 4206 & 4.465 & 0.5777& 370	(60)   & .09 & .0016\\ 
07592990 & 15.788 & 18.6 & 32.4$^\dagger$& 4004 & 4.632 & 0.4421& 500	(80)$^\ddagger$   & .05 & .0012\\
08314902 & 15.745 & 4.9 & 13.4$^\dagger$ & 4176 & 4.480 & 0.8135& 330 (50)/610 (100)$^\ddagger$ & .08 & .0010\\
10515986 & 15.592 & 9.7 & 18.1$^\dagger$ & 3668 & 4.297 & 0.7462& 350 (50)$^\ddagger$ 	   & .22 & .0030\\
11087527 & 15.603 & 6.5 & 9.1 		    & 4303 & 4.556 & 0.4110& 310 (70)/650 (300)      & .05 & .0012\\

\hline
12365719 & 15.843 & 7.6 & 19.4$^\dagger$& 3735 & 4.473 & 0.8501& inconclusive   &--&--\\
10063343 & 13.164 &&& 3976 & 4.433 & 0.3326& -- & 0.03 & --\\

\hline 
EY Dra 		 & &&&&&	0.4587 & 350 &\\ 
V405 And		 & &&&&&	0.4650 & 300 &\\
GSC 3377-0296 & &&&&&	0.4225  & 530 &\\
\hline

\end{tabular}
\caption{Basic data on the interesting \textit{Kepler} targets according to the Kepler Input Catalogue. Contamination values show minimum/maximum flux contamination from other stars,  $^\dagger$ shows targets, where light contamination change is higher than 8\%. The periods were found using automated discrete Fourier-transformation. The last two columns show our estimate for cycle lengths and the percentage of the period change.  L means a long-term trend in the plot. Kepler-10063343 is shown as an example, where no variations in the rotation period was found. Detected periods for targets marked with $^\ddagger$ is more uncertain, see Sect. \ref{sect:notes} for details. 
The last three lines show the objects from \protect\cite{shortcyc}, on whose the target selection of the current study was based.
The $\alpha$ parameter in the last column gives our lowest estimate of the differential rotation shear, see discussion in Sect. \ref{sect:discussion}.}

\label{tab:params}
\end{table*}

\begin{figure}
\centering
\includegraphics[angle=-90,width=0.47\textwidth]{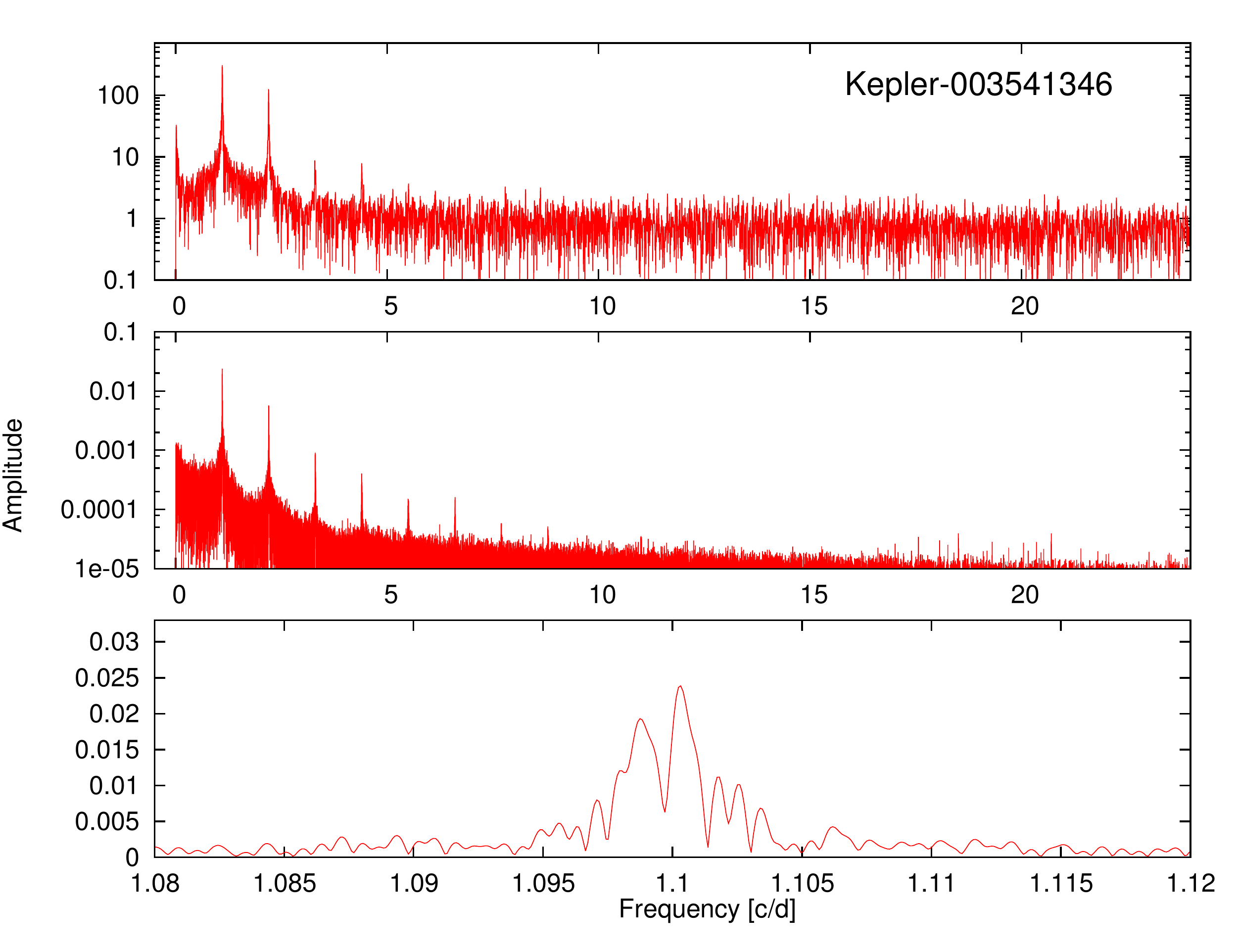}
\caption{An example Fourier-spectrum of Kepler-003541346. The spectrum on the top was made using unprocessed instrumental data (in flux units) only from Q1, used for the automated period search. The middle plot uses all available data, cleaned and interpolated, as used for further analysis (in magnitude units). The bottom plot shows the area of the rotation frequency, the multiple peaks indicate active regions with different rotation periods as a result of differential rotation. The upper two plots are in a logarithmic scale.}
\label{fig:fourier}
\end{figure}

 \cite{shortcyc} found activity cycles with periods on the order of one year on fast-rotating late-type stars, we based the target selection on these objects.
As first selection, we looked for dwarfs (with $\log g \approx 4.5$) cooler than 4500\,K in the Kepler Input Catalogue (KIC)\footnote{\url{http://archive.stsci.edu/kepler/kic.html}}, which resulted in 8826 objects. Then we analysed the light curves of these objects for one (Q1) quarter\footnote{
for the definition and dates of the Kepler observing quarters (marked as e.g. Q1) see \url{http://archive.stsci.edu/mast_faq.php?mission=KEPLER#35}
}
  of data using an automated method. This was done by applying discrete Fourier-transformation on the light curves using the one-dimensional option of {\it TiFrAn} (Time-Frequency Analysis) package\footnote{\url{http://www.konkoly.hu/tifran/}} \citep{tifran}. The highest peak of the resulting spectrum was then   assigned as rotational period of the given target (for an example of these spectra, see top plot of Fig. \ref{fig:fourier}). 
Note, that the highest peak in the Fourier spectrum does not necessarily correspond to the rotation period (e.g. if two active region are separated by $\approx180^\circ$, the double frequency can be stronger), but this method can be useful to select some suitable candidates from a large sample for more thorough inspection.
From the result of this automated Fourier analysis we selected 113 objects with short-period ($P_\mathrm{rot}\lesssim 1d$), then narrowed the list to single stars, where the light curve obviously indicated spottedness, i.e., the light variation showed continuous changes in time, resulting in 39 targets.  The formal error of the determined rotation periods can be estimated by using the frequency value at 90\% of the spectral window. This corresponds to 10\% precision (see \citealt{uzlib}). This is ~0.0025[1/d] in frequency, and is basically the same for all stars, as the properties of the data are the same. This formal error is however not quite meaningful, as the rotation period itself is changing in time.

For these objects all the publicly available long cadence data were downloaded from the \textit{Kepler} database (until Q16). 
For our analysis we used PDCSAP\verb+_+FLUX data created by the regular PDC-MAP algorithm, processed by pipeline version 9.0 (for more details see Kepler Data Release Notes\footnote{\url{http://keplerscience.arc.nasa.gov/Documentation.shtml}}).
We used light curves with version number 5.0, where the timing error of the \textit{Kepler} light curves is already corrected. For further investigation, the light curves were transformed from instrumental flux to magnitude scale (the small amplitude of the light curves causes negligible difference in the Fourier-frequencies resulting from the difference of the magnitude/intensity scale). 
 We fit each observing quarter by a third order polynomial, which was removed from the original light curves, giving a homogeneous light curve varying around zero magnitude. This helped us to get rid of the long-term instrumental changes.
These long-term trends might affect the Fourier-analysis, although the changes on this timescale should be independent in Fourier-space from the signals of the order of the rotation period.

As a sanity check we examined the contamination values provided by
the MAST catalog\footnote{\url{http://archive.stsci.edu/kepler}}  for those targets, that indicated possible activity cycles (see Sect. \ref{sect:results}). In each 
case we found a non-zero value: the numbers range from a few percent
up to 25-30\% (depending on the observing quarter, see Table \ref{tab:params}). The percentage refers   
to an estimate of the flux that comes from the contamination, i.e., close-by 
stars contributing to the sum of flux in the assigned Kepler aperture mask.

These relatively high numbers motivated a more thorough check, namely we
downloaded the target pixel files for our 10 targets and checked whether
the rotation signal comes from the target or a close neighbor. To do this 
we choose the most contaminated quarters for each stars. We found that in    
all cases the the rotational variation originates from the target star and 
within Kepler's resolution no additional variation is found from off-target
pixels. We have to mention that although a long-term contaminating variation 
from nearby stars (that might alter the amplitude of the rotational variation) 
would be much harder to exclude this way, our target stars are much more brighter 
than their surrounding pixels, so a large amplitude variation that would
compromise the conclusions of this paper can be safely exclude in all cases. 

\section{Method}
\label{sect:method}

\begin{figure*}
\centering
\includegraphics[width=0.32\textwidth]{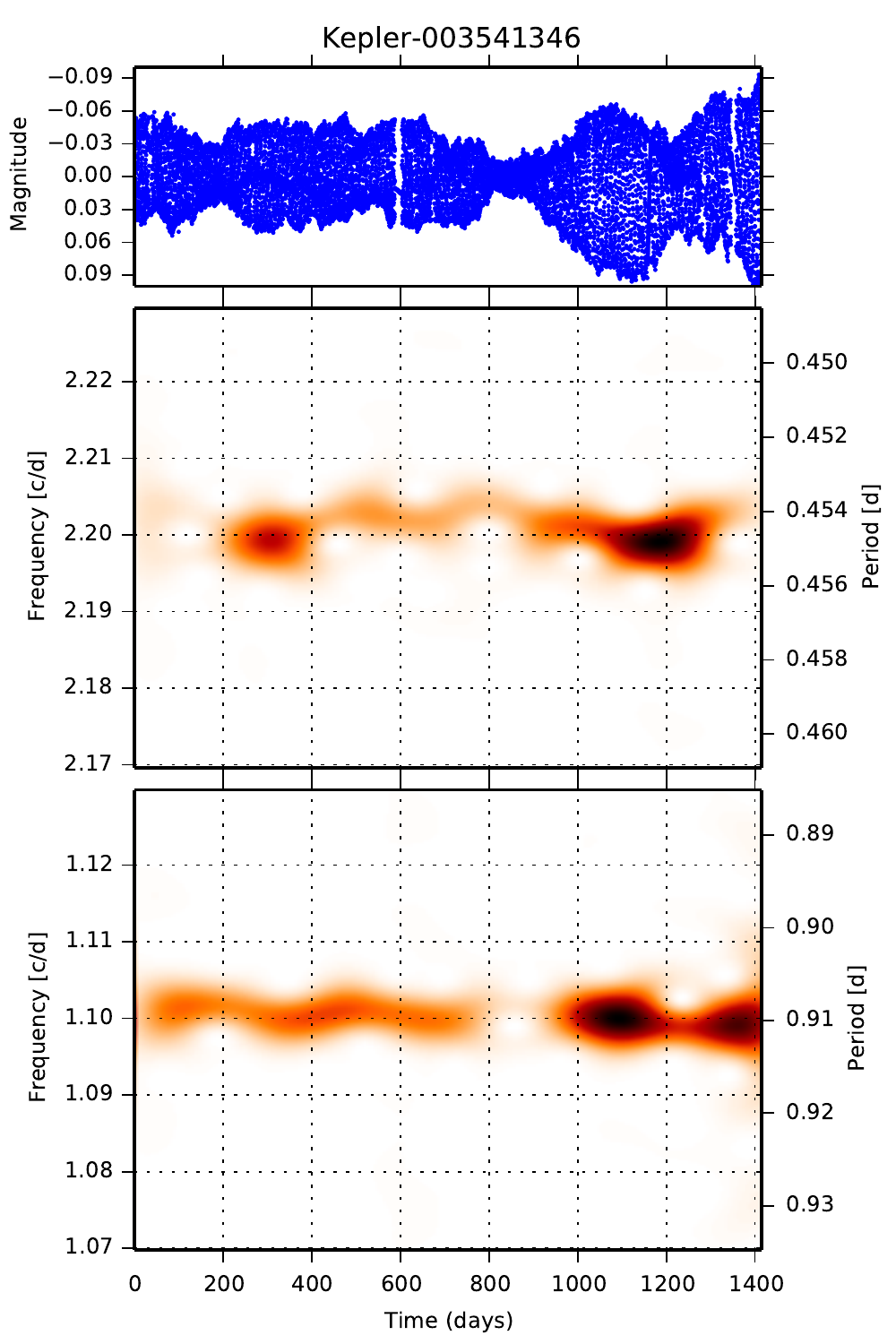}
\includegraphics[width=0.32\textwidth]{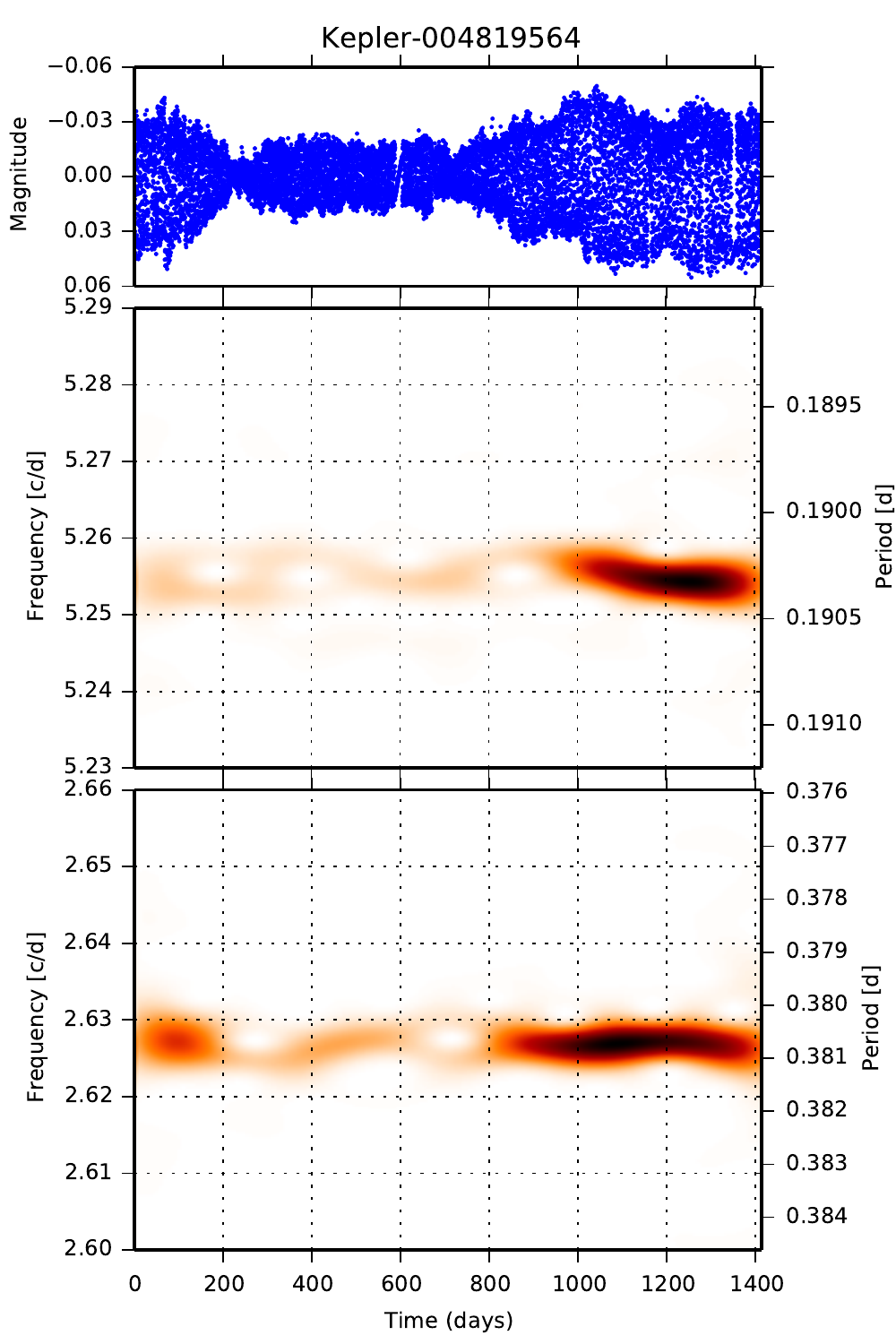}
\includegraphics[width=0.32\textwidth]{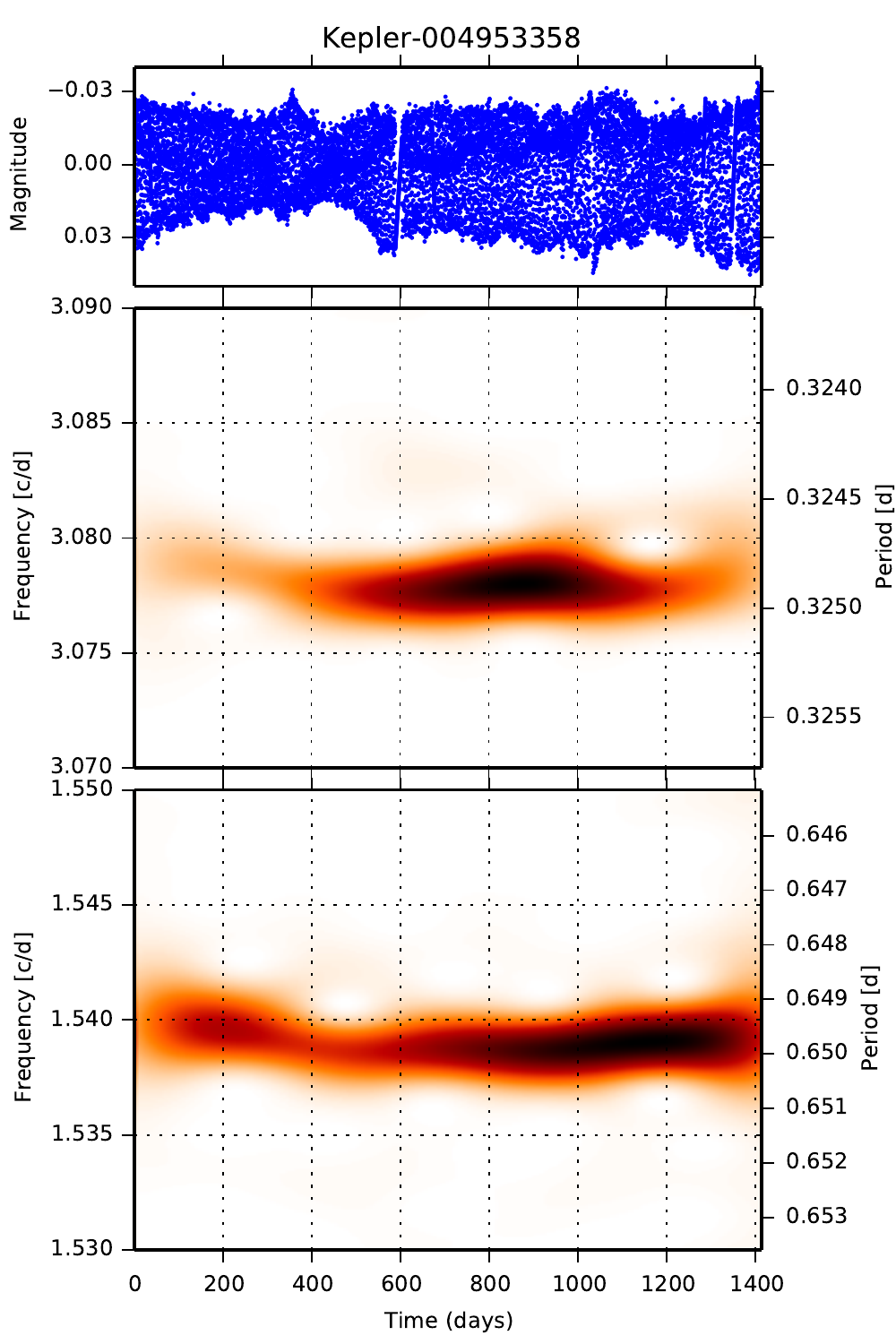}
\includegraphics[width=0.32\textwidth]{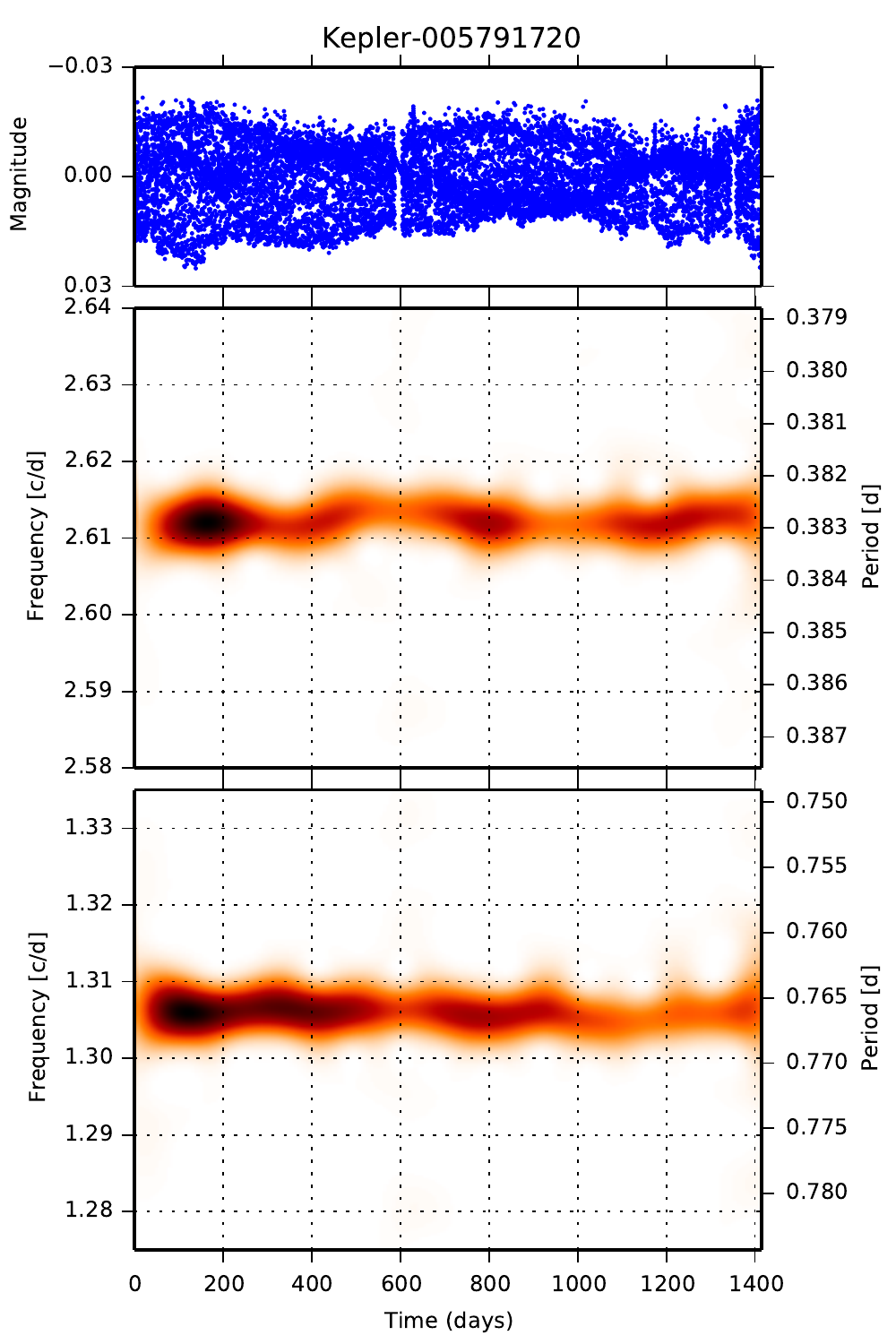}
\includegraphics[width=0.32\textwidth]{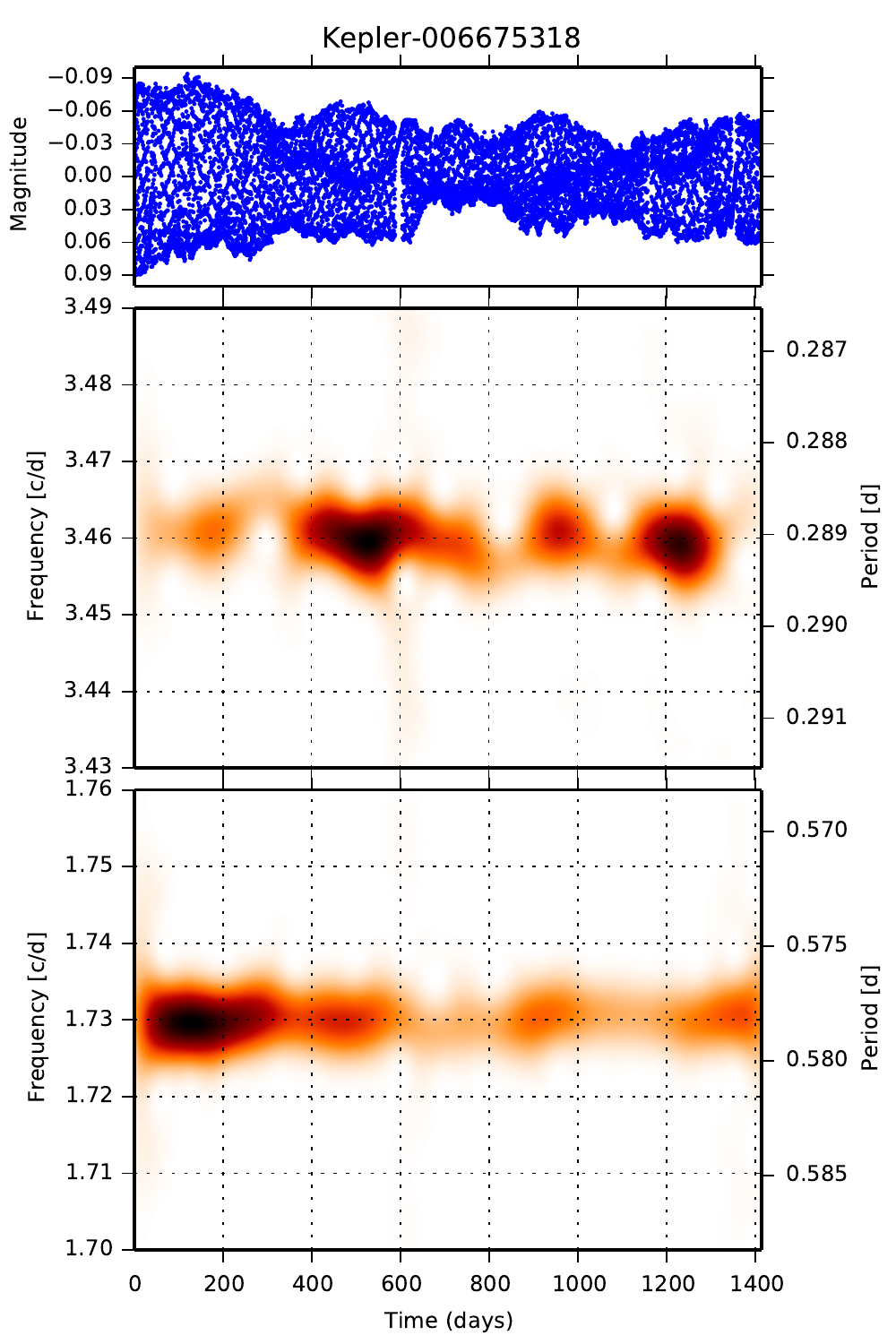}
\includegraphics[width=0.32\textwidth]{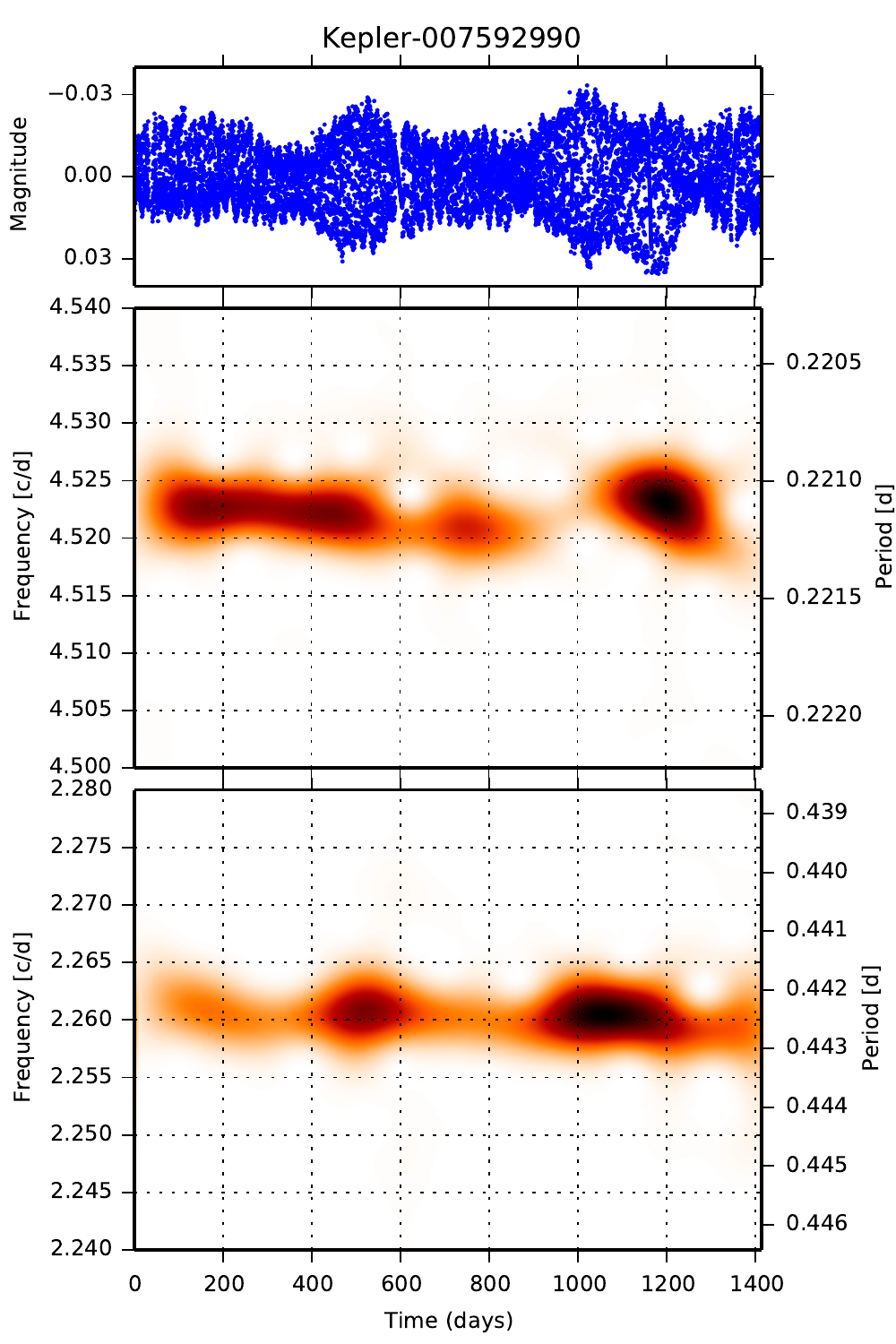}
\caption{Cleaned and interpolated light curves (top panels), and their short-term Fourier-transforms (middle and bottom panels) for the \textit{Kepler} targets where sign of activity cycles was found. The middle and bottom panels show the STFTs at the double of the rotation frequency, and  at the rotation frequency, respectively. Kepler-10063343 is shown as a comparison, where no change in the rotation frequency was found.}
\label{fig:stft}
\end{figure*}

\begin{figure*}
\centering
\includegraphics[width=0.32\textwidth]{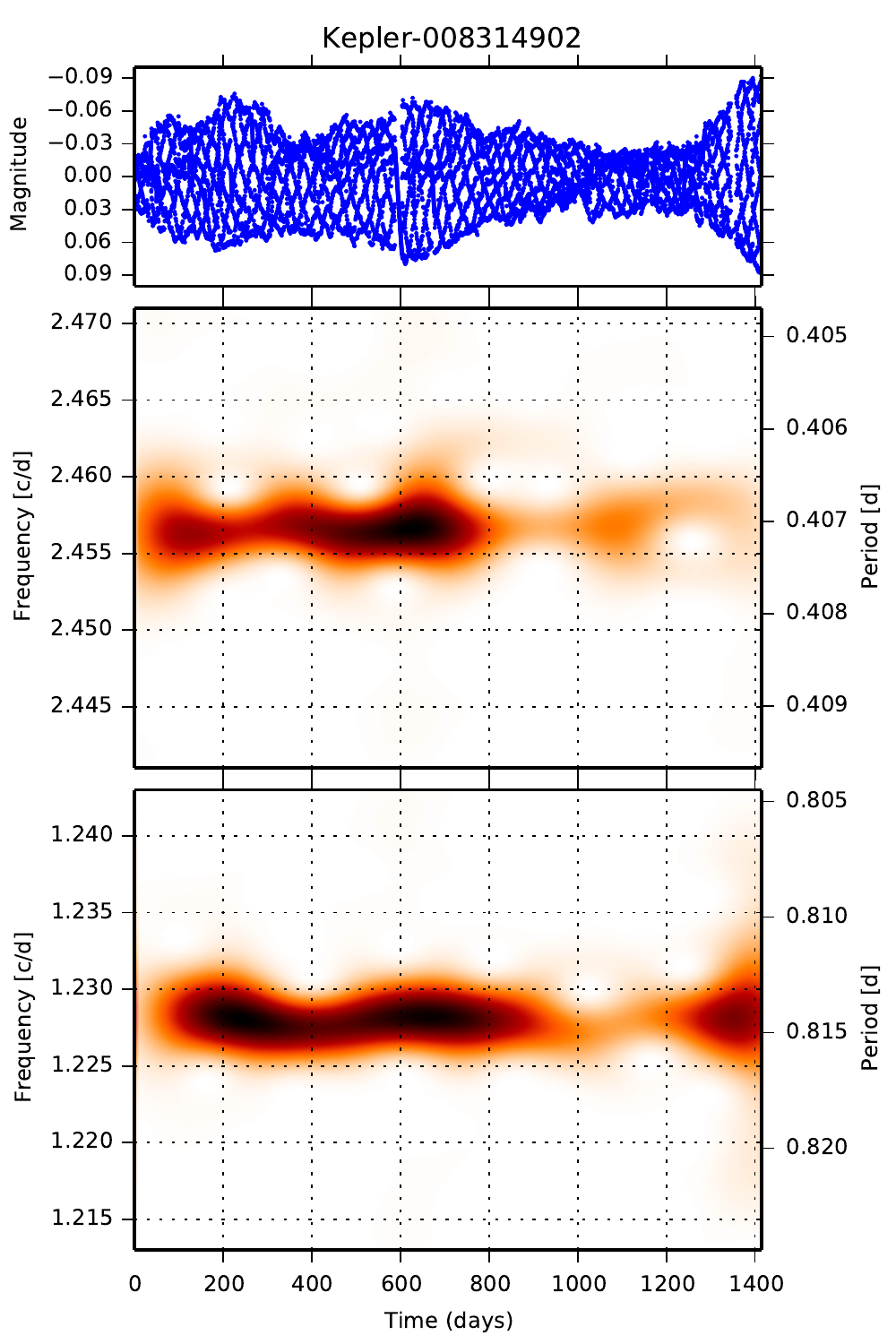}
\includegraphics[width=0.32\textwidth]{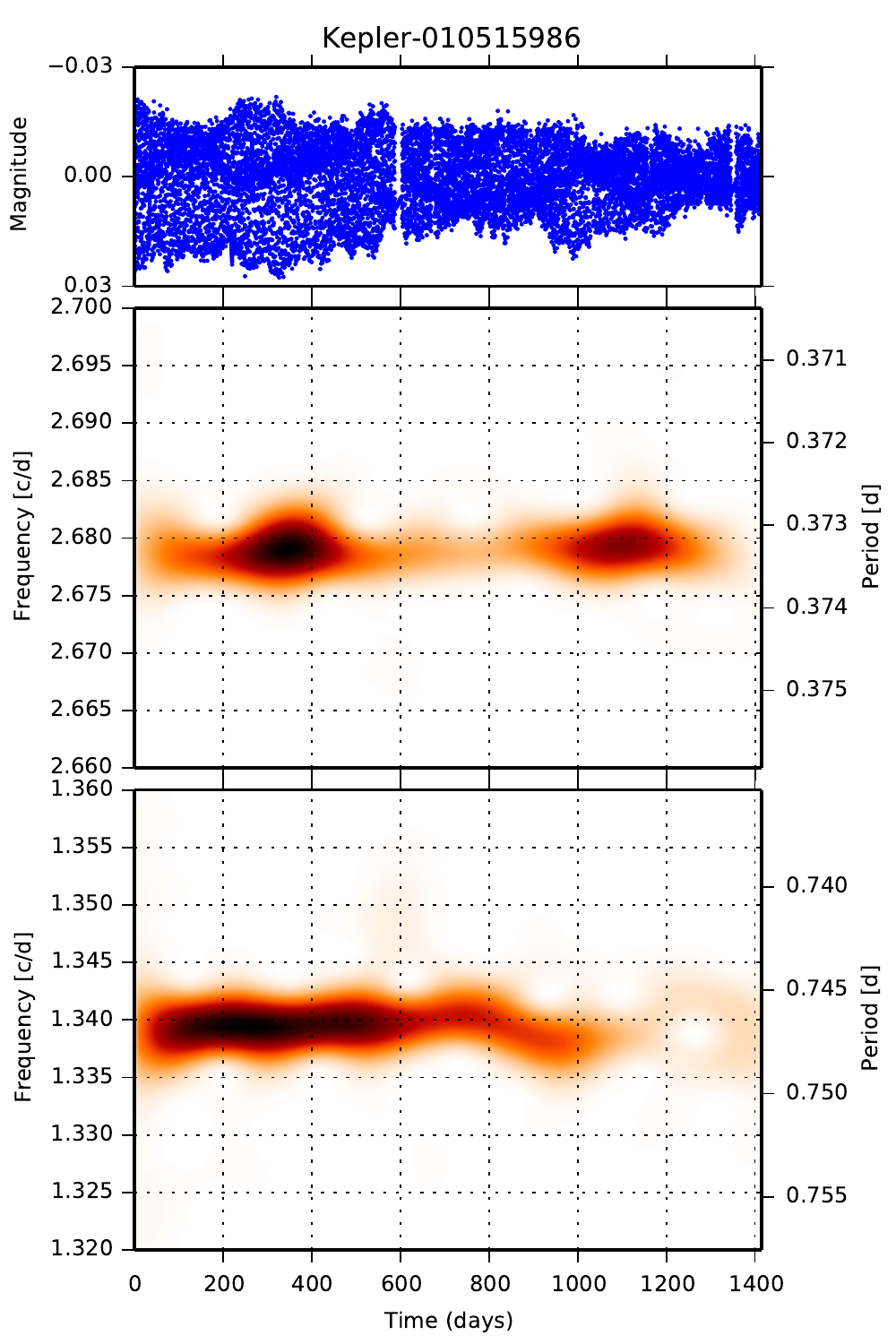}
\includegraphics[width=0.32\textwidth]{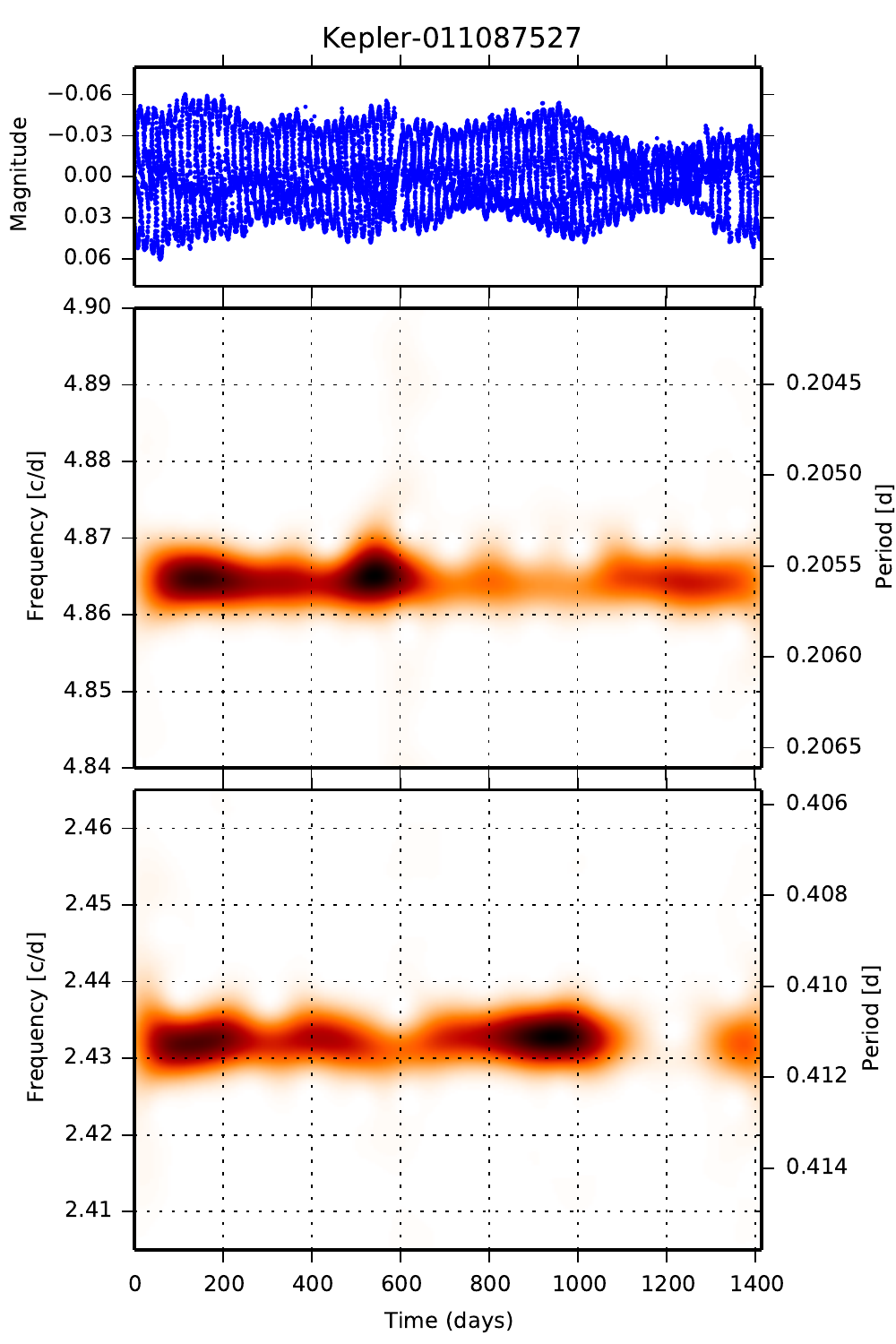}
\includegraphics[width=0.32\textwidth]{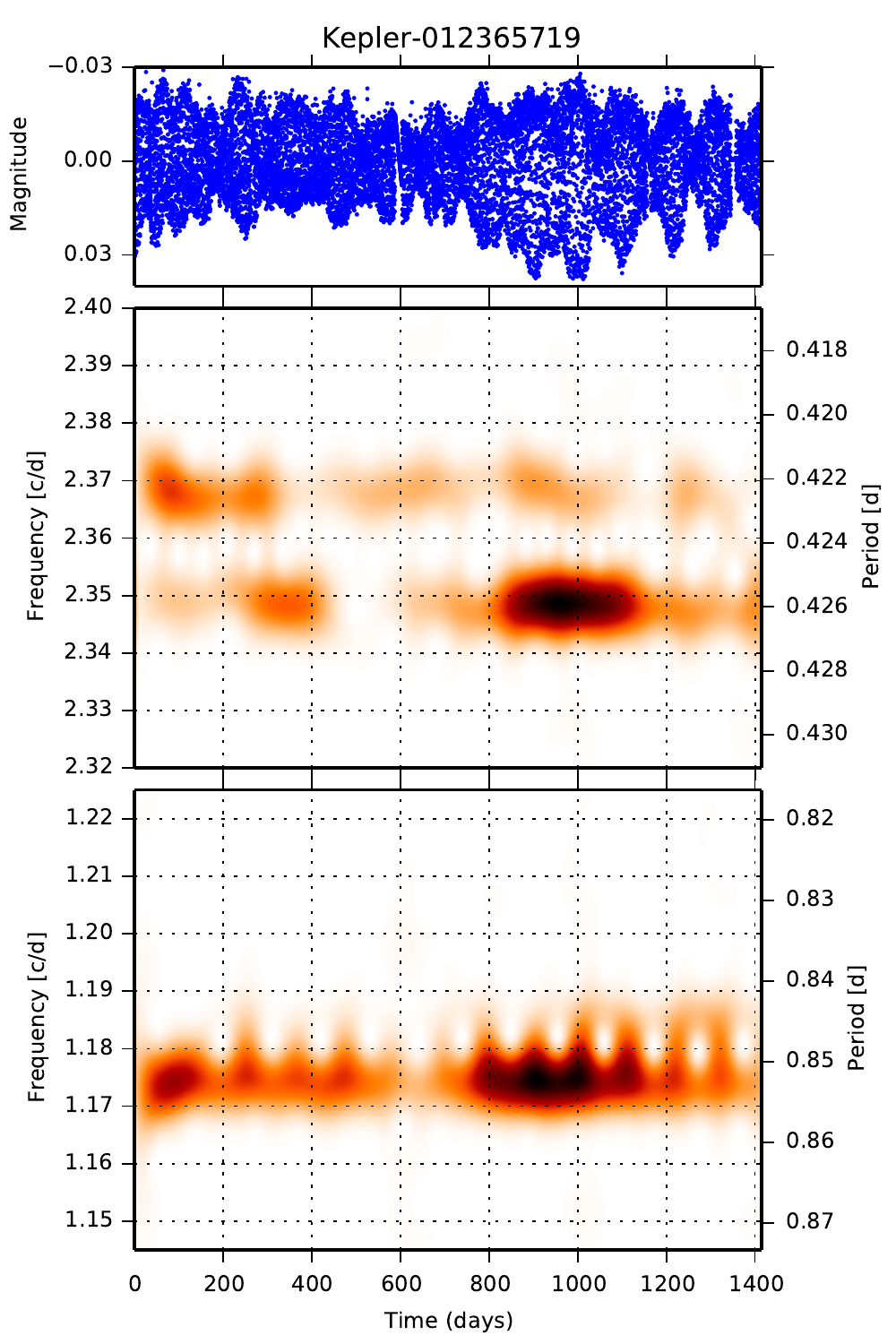}
\includegraphics[width=0.32\textwidth]{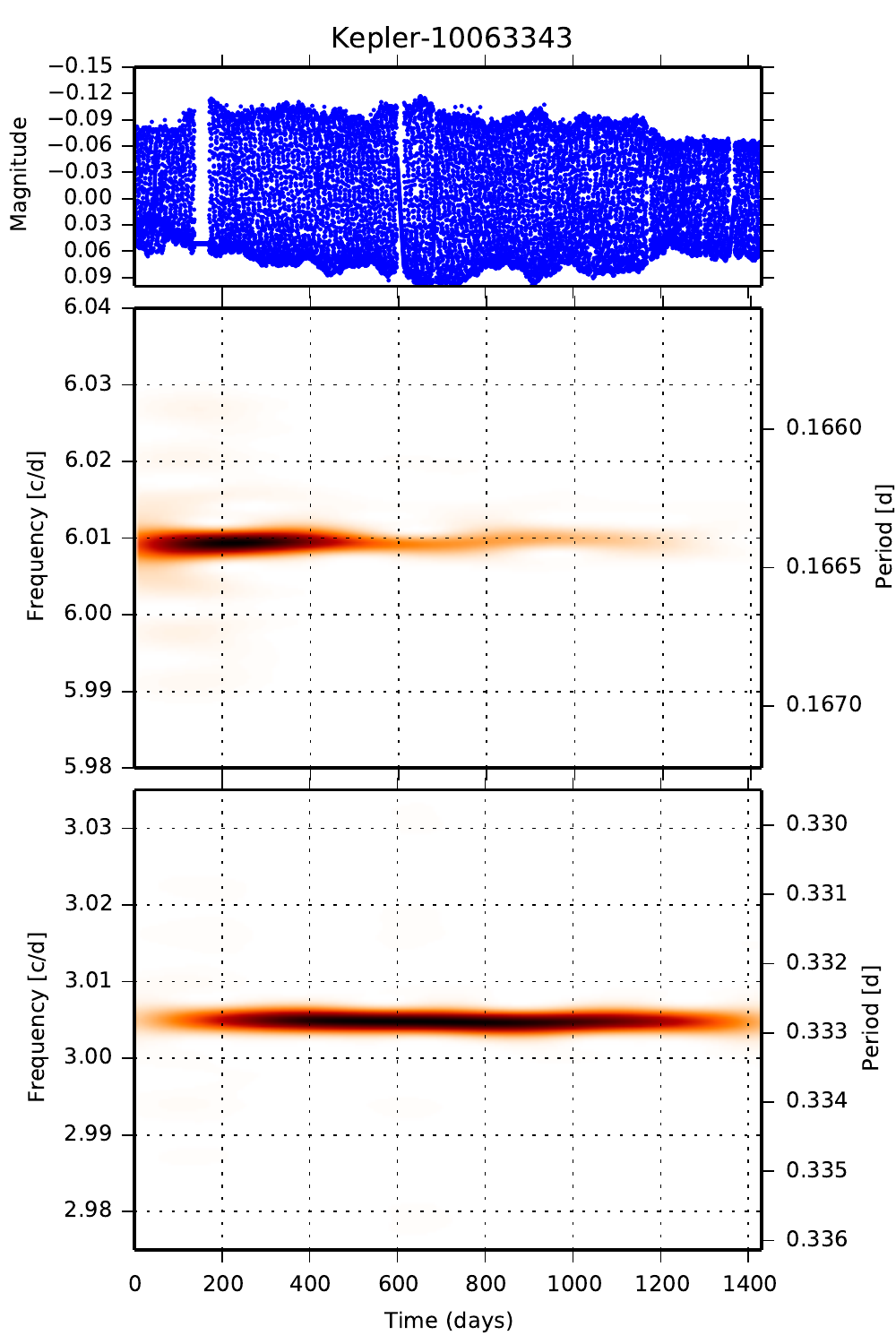}

\contcaption{}
\end{figure*}

\begin{figure}
\includegraphics[width=0.47\textwidth]{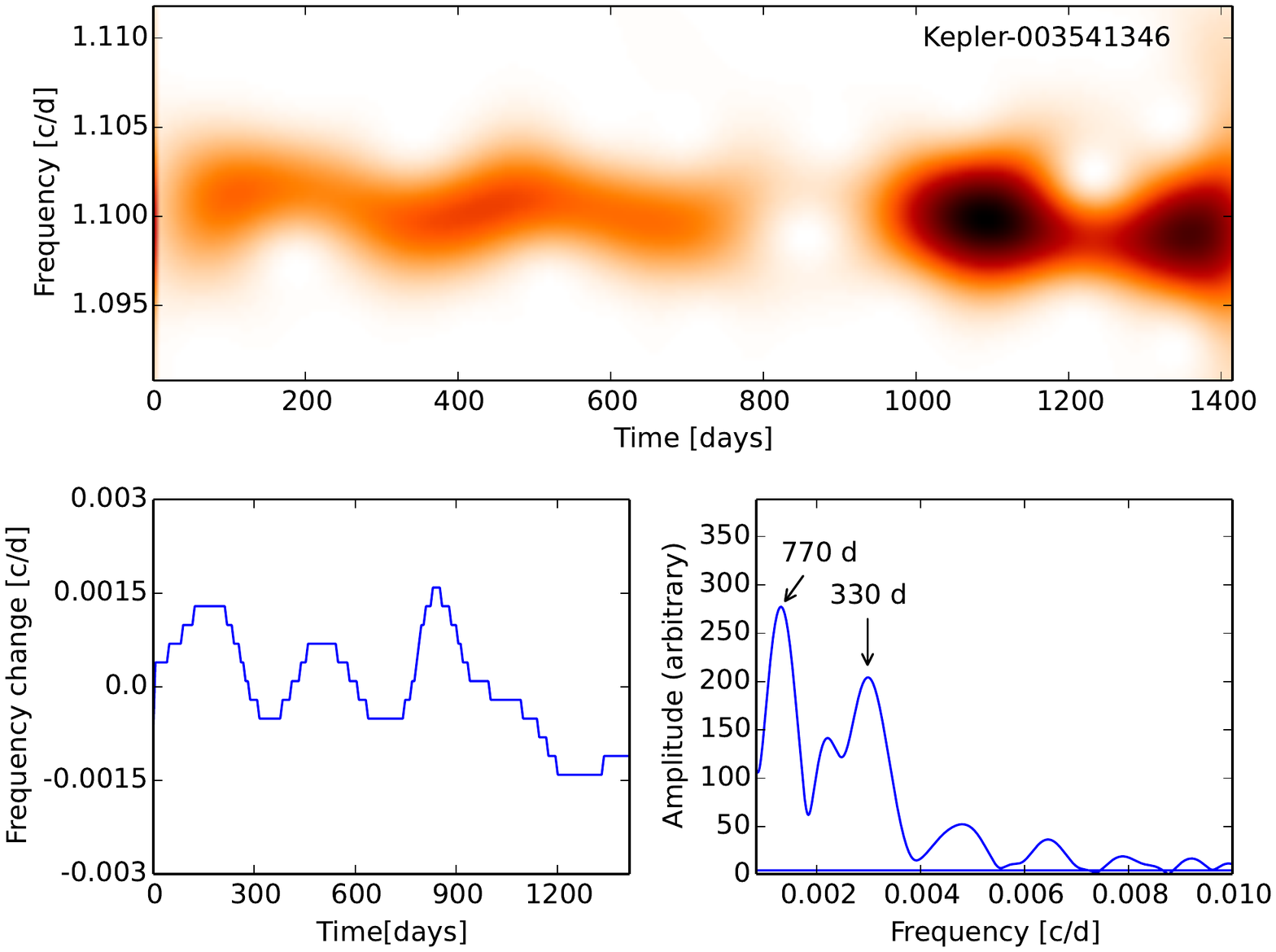}
\caption{\textit{Top:} STFT of the Kepler-003541346 light curve, \textit{bottom left:}  detected frequency change in the STFT, \textit{bottom right:} Fourier-transform of this change with a (probably fake) signal at 770 days, and a real signal at 330 days.}
\label{fig:python}
\end{figure}

The data were examined using various approaches of the {\it TiFrAn} package to select the promising candidates. The light curves of the selected targets were cleaned from the extremely outlying points, caused by flares. According to \cite{2009AA...501..695K}, Fourier-signals of ill-sampled data are much harder to recover, thus the light curves were interpolated to the \textit{Kepler} sampling to cover the gaps in the observations. We used linear interpolation to fill the gaps in the datasets, which results in straight lines in the missing parts, but this does not alter the main results. We used Short-Term Fourier-Transform (STFT) in this study and report the results in Fig. \ref{fig:stft}; for a description of this and other, different approaches see \citealt{2009AA...501..695K}. 

An STFT plot might look incomprehensible at first sight, but it is a most helpful tool for time--frequency analysis. The STFT method applies a Gaussian window on the light curve, and by moving this window a sequence of Fourier-spectra is obtained. The vertical changes in these plots indicate a shift in a peak in the Fourier-spectrum in time,  the z-axis (color code) shows the amplitude of the signal.
An $\alpha$ parameter (see Eq. 2 in \citealt{2009AA...501..695K}) regulates the width of the Gaussian window, i.e., the balance between temporal and frequency resolution. This was fine-tuned for each cycling target in Fig. \ref{fig:stft} to recognize the changes easier.

Strong light contamination from other sources in the {\it Kepler} aperture can decrease the amplitude of the light curves. This effect can decrease the precision of the frequency determination. Thus, such contamination could make the determination of the cycle lengths harder (even more if it changes through observing quarters), but it does not change the frequency itself. This effect might be seen in Table~\ref{tab:params}: targets with higher light contamination change (marked with~$^\dagger$) have less confidently determined activity cycles (marked with~$^\ddagger$). The tests on artificial data however showed, that the method we used to determine the cycle lengths is not very sensitive to amplitude changes (see Appendix \ref{sect:appendix}).

\section{Results}
\label{sect:results}

In 9 of the 39 promising targets we found signs of cyclic or long-term modulations in the rotation period, which we interpret as a  manifestation of stellar activity cycles. The basic parameters of these targets are summarized in Table \ref{tab:params}. The light curves and the time--frequency analysis of these objects is plotted in Fig. \ref{fig:stft}. The last plot in Fig. \ref{fig:stft} is used as comparison, for that star, which otherwise is similar to the studied objects, no such modulation was found. 

We know from the Sun, and also from long-term photometric analysis of active stars \citep{2009AA...501..703O} that activity cycles are not strictly regular phenomena. Multiple cycles exist, and also the length of the cycles change in time. Therefore, one should be cautious when trying to fit exact parameters to quantify the results, often a mere visual inspection can be the most effective, especially, since the datasets cover only a few cycles.

 However, as an experiment for a quantitative description, we analysed the STFT of the light curves using a method based on the Fourier-transformation of the STFT maxima  around the rotation frequency: basically we were looking for detectable periodic signals in the STFT itself (see Fig. \ref{fig:python}).
This method, however, should be used with much caution and sanity, as signals with periods in the order of the length of the datasets are often present and are not necessarily real.  Also, if light curve amplitudes change very much from cycle to cycle the recovery of cycle length by this correlation method could get difficult. A clear estimate of the cycle timescale directly from the STFT helps to find the real signal in this Fourier-spectrum  (as human brain is much better in pattern recognition than any software). The result of this analysis is summarised in Table \ref{tab:params}. This method can be also used to estimate the uncertainties of the cycle lengths. Using the frequency difference between the maximum and half-maximum of the peak in this Fourier-transform as an error gives 50--100 days as the typical uncertainty of the cycle lengths.

We analyzed each target using discrete Fourier-transformation with MuFrAn  (Multiple Frequency Analysis, \citealt{1990KOTN....1....1K}) as well, examining 30--200 day-long subsets of the light curve. This way the stability of the rotational periods of the active regions present in that segment could be checked. 
 As starspots appear on the surface at different latitudes, multiple peaks appear in the Fourier-spectrum, thanks to the differential rotating surface (see bottom panel of Fig. \ref{fig:fourier}). 
In case of continuous period change due to differential rotation, the length of the light curve segment shows that in how long dataset the double or multiple peaks appear in the Fourier-spectrum, signaling the measurable change of the rotational period of the star. 
 As the typical spot emergence latitude -- the activity belt -- is shifting with the butterfly diagram, the observed main rotation frequency is also changing. For the same reason, the $2f$ regions can have different structure from the region of the main frequency: if spots are separated by $\approx180^\circ$, but they are at slightly different latitudes, their signal in the Fourier-spectrum can be stronger than the signal at exactly the $2f$ frequency.

 \section{Notes on individual objects }
 \label{sect:notes}
 
\begin{description}

\item[Kepler-03541346] The signal of the rotation changes smoothly in time, an activity cycle with a period of $\approx330$ days seems to exist. The cycle can be seen pronounced at the rotation frequency ($f$), and a similar trend is visible at its double ($2f$), especially between 400--900 Kepler-days\footnote{starting at JD 2454833.0}, meaning that the shape of the cycle is not symmetric. 

\item[Kepler-04819564]  We find a  periodicity of $\approx530$ days. Frequency splits can be observed in both the rotational frequency $f$ and its double $2f$, indicating two active regions present on different latitudes with slightly different rotation periods.

\item[Kepler-04953358] A longer-term variability is seen in the rotation frequency and its double frequency runs parallel with a  period of $\approx600$ days.  The long-term variation dominates the time--frequency plot, but to confirm its existence, longer dataset is needed.

\item[Kepler-05791720] We find a  periodicity of $\approx320$ days, which is present in both the $f$ and $2f$ frequencies. The light curve shows very frequent flare activity which is seen in Fig. \ref{fig:stft} as deviating brighter data points.

\item[Kepler-06675318] A hint for a periodicity  of $\approx370$ days can be seen. The $2f$ plot indicates a periodicity of similar value. 

\item[Kepler-07592990] There is  a sign for a $\approx500$ day-long variation in the rotation frequency. Beside the rotation signal, the Fourier-spectrum shows another period around 20 days. 
According to DSS\footnote{ Digitized Sky Survey, available at \url{http://aladin.u-strasbg.fr/}} images there is a close-by source to this target. Thus, we examined the target pixel file from Q12, to find out the origin of this signal. We checked Fourier-transforms of light curves from different pixels and pixel sets. According to this analysis, this longer, 20 day-long period is most possibly associated with the contaminating source. Unfortunately, the two objects are too close to each other to create a contamination-free aperture. We checked the STFT analysis after prewhitening the light curve with the 20 day-long period, there was no change in the result.

\item[Kepler-08314902] In the region of the rotation frequency  a variation with a period of $\approx610$ days is seen.    The region of the double frequency shows a  clear variation of $\approx$330 days, and a period of $\approx470$ days can also be seen. These double signals might be a result of multiple cycles, but longer dataset is needed to confirm.

\item[Kepler-10515986] Certainly there is a variation in the rotational period, but no typical cycle length can be easily given, the first 500 days show quite irregular behavior.  Our best estimation is 300--400 days. The light curve shows frequent flares.

\item[Kepler-11087527]  The $f$ region suggests periodicites of $\approx$310 and $\approx650$ days, that are also present in the $2f$ region.

\item[Kepler-12365719] Two distinct frequencies can be seen on this target, and there is {a variation of in the rotational frequency with a time scale of a few hundred days. The two nearby frequencies does not allow us to determine a more precise value.} We examined both DSS images, and the \textit{Kepler} Full Frame Images, but no visible close-by stars could be seen. Since the pixel size of the  \textit{Kepler} CCD is 4", a contaminating source cannot be safely ruled out by visual inspection of the images. Another interesting possibility could be that this target is a close binary system with components having slightly different rotational periods, similar to BY Dra \citep{Pettersen1992tg}. If this is a single target, that would mean that two very persistent active nests are present on the object, at two distinct latitudes during the full length of the observations which do not move, and no (or undetectable) other active region appears. We consider this last scenario less likely.

\end{description}

\section{Discussion}\label{sect:discussion}

\begin{figure}
\centering
\includegraphics[width=0.45\textwidth, bb=50 350 554 800]{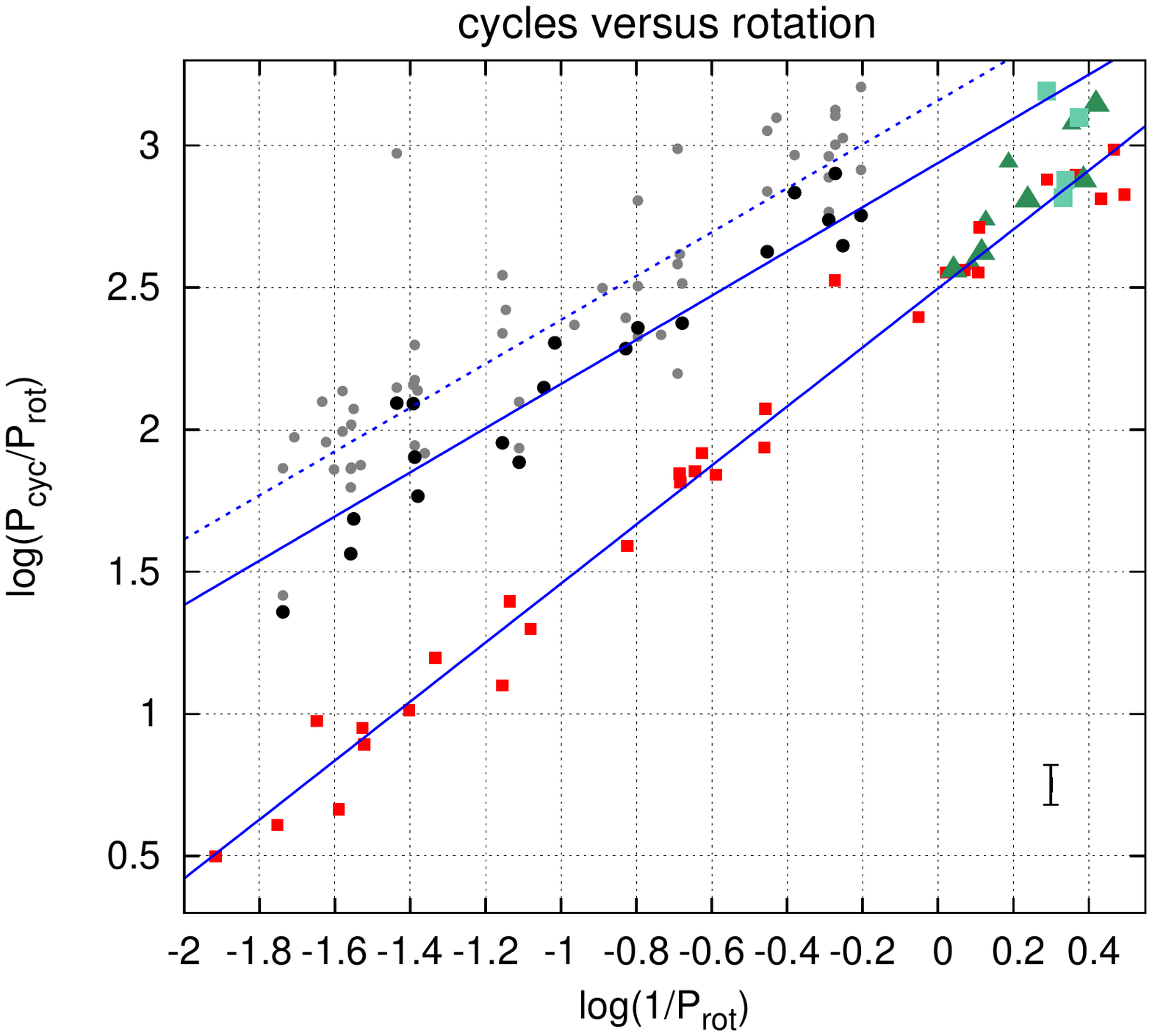}
\caption{Correlation between the rotation period and the length of the activity cycle, as in \protect\cite{2009AA...501..703O}. Large black dots,  green squares and  green triangles stand for the shortest cycles from \protect\cite{2009AA...501..703O}, results from \protect\cite{shortcyc}, and from the present paper, respectively. Smaller grey dots denote data from different surveys, from \protect\cite{2009AA...501..703O}.  Smaller triangles mark the less certain periods from this paper (marked with $^\ddagger$ in Table \ref{tab:params}). Red filled squares show data for M dwarf stars from \protect\cite{savanov}. The dotted line represents the fit to all the data from \protect\cite{2009AA...501..703O}, \protect\cite{shortcyc}, plus the results of the present paper excluding M stars; while the parallel line shows the fit to the shortest cycles of that dataset. The slope of the fit to the data from  \protect\cite{savanov} is close to 1.0, which means that no correlation is found between rotation and cycle lengths.   Error bars on the lower right indicate the typical uncertainty of rotation period and activity cycle determination (error bar of the x axis is smaller than the line itself). See the text for more.}
\label{fig:rot-cyc}
\end{figure}

\cite{keplerrot} studied the rotation of about 2500 {\it Kepler} M dwarfs with a method based on autocorrelation. Their sample with detected rotation contains four of our cycling targets: KIC 4953358, KIC 5791720, KIC 10515986, and KIC 12365719. In all cases their rotation periods derived by an independent method shows very good agreement with our values.

\subsection{Estimation of the differential rotation shear}
We estimate the $\alpha$ parameter, which is often used in practice to describe the shear of the differential rotation, and is defined as 
$\alpha= \Delta\Omega/\Omega_\mathrm{eq} = (\Omega_\mathrm{eq}-\Omega_\mathrm{pole} )/\Omega_\mathrm{eq}$. 
If we suppose that the extrema of the rotation periods ($P_\mathrm{rot,min}$, $P_\mathrm{rot,max}$) we found represent the rotation at the equator and the pole, we can give a lowest estimation of the $\alpha$ parameter using the $P_\mathrm{rot}$ and $\Delta P_\mathrm{rot}$ for each object -- these values are summarized in Table \ref{tab:params}. Note, that by using this method, it is not possible to determine if the differential rotation is solar or anti-solar, since we have no latitude information from where the rotational signal originates.

It is possible that the actual $P_\mathrm{rot}$ values span a wider range than we estimate from the extreme positions of the rotation frequency in the STFT diagram -- this can be revealed by detailed Fourier-analysis of segmented light curves for each target, but that is outside the scope of the current paper. This would  yield higher values of $\alpha$. 
A similar method was used by \cite{keplerdr} to study rotation and differential rotation in more than 40,000 active {\it Kepler} targets. The authors used Lomb--Scargle periodograms to find the extrema of the rotation periods during one quarter (Q3). 
Their sample contains seven targets from this paper\footnote{
KIC 03541346, KIC 04953358, KIC 05791720, KIC 06675318, KIC 08314902, KIC 10515986, and KIC 12365719
},
and in two cases they also detected sign of differential rotation: indeed, they found higher values of $\alpha$, whereas their main rotation periods for the matching targets agree with our values. In the case of KIC 3541346 and KIC 4953358 they found $\alpha=0.0101$, and $\alpha=0.0123$, respectively. The authors did not (and could not) take into account that the active regions might emerge only in a smaller latitude range, thus the differential rotation shear could be even higher. Given the nature of the method, the uncertainty in the spot latitudes does not allow accurate determination of the shear.
Note, that the authors used an oversampling of 20 that might result in peaks that are not real, and could change the values of $\alpha$ they found.

To give a better estimate, we should know the actual latitude values where the active regions emerge during the cycle, for which we can give only crude guesses. By assuming the usual quadratic differential rotation law of 
\begin{equation}
\Omega(\vartheta)=\Omega_\mathrm{eq}(1-\alpha\sin^2\vartheta),
\end{equation}
the shear can be determined from known rotation periods at given latitudes using the following equation: 

\begin{equation}
\alpha=\frac
{\Omega(\vartheta_2)-\Omega(\vartheta_1)}
{\Omega(\vartheta_2)\sin^2\vartheta_1 - \Omega(\vartheta_1)\sin^2\vartheta_2}.
\end{equation}

If we assume latitudes similar to the solar case, where the spots emerge between $0^\circ$ and $30^\circ$ latitudes, the given $\alpha$ values are higher by a factor of $\approx4$, however this scenario is unlikely (see e.g. \citealt{emre}). In a case based on the fast-rotating ($P_\mathrm{rot}=2d$) K0 dwarf model of \cite{emre}, where the spots emerge between $\approx 35^\circ$ and $\approx 45^\circ$ latitudes (cf. Fig. 10 of that paper), the $\alpha$~values in Table \ref{tab:params} get higher by a factor of $\approx6$ and we get a typical value of $\alpha=0.010$ for our sample.

For a more realistic estimate a dynamo model for fast-rotating late-type dwarfs is needed that could more accurately predict the emergence latitudes. 
\cite{magneticwreath} studied MHD models of a fast rotating Sun, and found wreath-like magnetic structures in the convection zone around 5--25$^\circ$ latitudes. \cite{emre} presented models of solar-like and main sequence K-type stars of different rotation rates, but a model of ultrafast-rotating late-type stars is a real challenge for the theoreticians.
Another way for getting more accurate values for $\alpha$ would be to determine directly the active region latitudes from the light curves by analytic modeling or inversion of the light curves. However, the information on the actual latitudes is very limited in photometric data, and is present only in the limb darkening of the spotted surface.


\subsection{Rotation--cycle length relation}

In \cite{2009AA...501..703O} a new correlation between the cycle length normalized with the rotation, and the inverse rotation period is given in log-log scale, with one single M dwarf star in the sample (EY Dra) with a quite short cycle period, which was not used in determining the slope of the relation. \cite{shortcyc} find that the activity cycles for ultrafast-rotating dwarfs are somewhat shorter than the previous samples would indicate, by extending the relation based on stars of slower rotation. In that work another M dwarf, V405 And, a binary, was added to the stars with known cycles.

Using the activity cycles from \cite{2009AA...501..703O}, \cite{shortcyc}, and adding the results of the present paper we studied again the rotation--cycle length relation (see Fig. \ref{fig:rot-cyc}). Two stars from the present sample seem to have double cycles, but the longer ones are uncertain due to the limited length of the dataset, thus only the shortest cycles have been considered. \cite{savanov}, using data from the ASAS survey for a homogeneous set of only M dwarf stars, dis not find any relation between the lengths of rotations and cycles.
Looking at Fig. \ref{fig:rot-cyc} we find, that four stars with certain cycles derived in the present paper, and two stars from \protect\cite{shortcyc}, V405 And and EY Dra, (which are M dwarfs), fit well the M dwarf sequence by \cite{savanov}, except one star, KIC 04819564. We thus excluded the M dwarf stars from the fits of Fig. \ref{fig:rot-cyc} but included KIC 04819564. The slope for all the cycles and for the shortest cycles (in case of multiple cycles) is $0.77\pm0.06$ and $0.78\pm0.05$, similarly to the earlier values of  0.74 by \citealt{1996ApJ...460..848B}) and to 0.81 for all the data and 0.84 for the shortest cycles by \cite{2009AA...501..703O}. 

The existence of a relation between the cycle lengths and rotational periods for a diverse sample of active stars (the sample contains both single and binary stars, giants and dwarfs, of different spectral types) did not change with the exclusion of the M dwarf stars. Already the short period part ($P_\mathrm{rot}<1$\,day) of cycling stars seems to separate by spectral type, and was not evident in \protect\cite{shortcyc}  which was prepared before the results of  \cite{savanov} was published, without the table cycle lengths.

The relation between rotational and cycle periods is extremely important for understanding stellar dynamos (see e.g. \citealt{1996ApJ...460..848B,dynamo2,magneticwreath, dynamo1} for the details). The results presented in this paper, which populate the short period end of the rotation--cycle length relation, give a good impact to this study, showing a clear separation between the K and M dwarfs already at very short rotational periods. The determination of the exact spectral types of cycling stars studied in this work is the subject of a forthcoming paper.

\section{Summary}
\label{sect:summary}

\begin{itemize}
\item We analyzed light curves of 39 fast-rotating ($P_\mathrm{rot}\lesssim1d$) active stars from the \textit{Kepler} database using time--frequency analysis.

\item From the short-term Fourier-transforms (STFT) of the light curves in the region of the rotation frequency and its double, we detected quasi-periodic variations.

\item We interpret these variations as a result of stellar butterfly diagram: during the activity cycle the typical latitude of the starspots change, and this, because of the differential rotation of the surface, results in change of the rotation period.

\item  With our technique, we found hints of activity cycles with periods in the range of 300--900 days in 9 targets.

\item To find activity cycles through rotational period variation due to differential rotation and the butterfly diagram is a new method which does not need latitudinal information as input, applicable only for very high precision and (nearly) continuous datasets such as is produced by the Kepler satellite. 

\item This result populate the short-period part of the rotation--cycle length relation   showing clear separation between the K and M dwarf stars with the shortest periods($P_\mathrm{rot}<1$\,day), this is very important in understanding the nature of the cycling dynamos. 

\end{itemize}

\section*{Acknowledgments}
We would like to acknowledge the helpful comments of the anonymous referee, which improved this paper considerably.
We thank Gy. Szab\'o
for the help with \textit{Kepler} data and M. V\'aradi for his suggestions with the period analysis. The authors are grateful for I. Savanov for the table of M-dwarfs  cycles. We would like to thank the anonymous referee for the helpful comments, which improved the paper significantly.
The financial support of the OTKA grants K-81421, K-109276, K-83790, the KTIA URKUT\_10-1-2011-0019 grant, and  the European Community's Seventh Framework Programme (FP7/2007-2013) under grant agreements no. 269194 (IRSES/ASK) and no. 312844 (SPACEINN) is acknowledged.
This work was also supported by the ``Lend\"ulet-2009'', and ``Lend\"ulet-2012'' Young Researchers' Programs of the Hungarian Academy of Sciences, and by the HUMAN MB08C 81013 grant of the MAG Zrt.
R.Sz. was supported by the J\'anos Bolyai Research Scholarship of the 
Hungarian Academy of Sciences.
Funding for the \textit{Kepler} mission is provided by NASA's 
Science Mission Directorate.

\appendix
\section{Testing the STFT to find butterfly diagrams on solar and artificial data}
\label{sect:appendix}

\begin{figure*}
\centering
\includegraphics[width=0.45\textwidth]{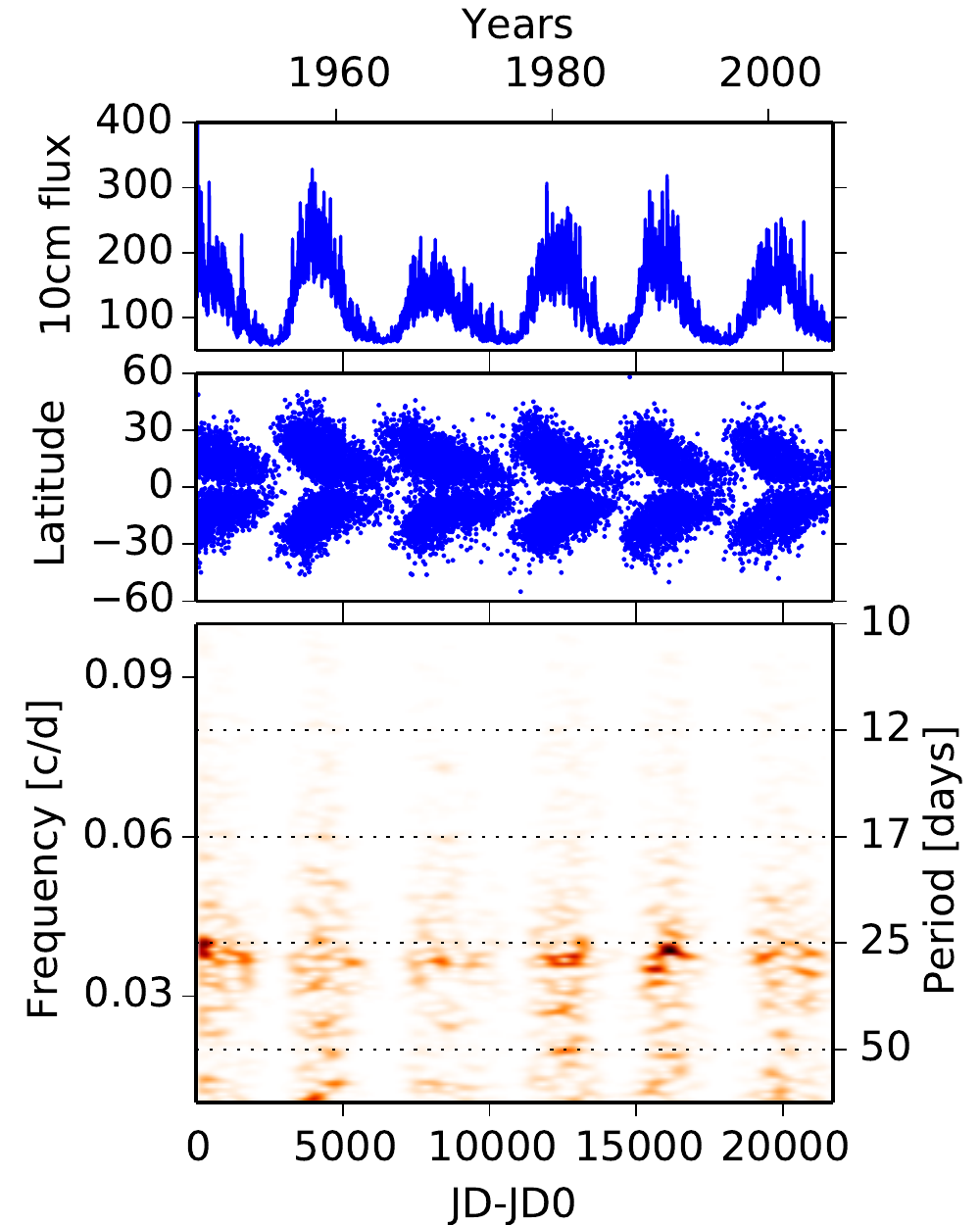}
\includegraphics[width=0.45\textwidth]{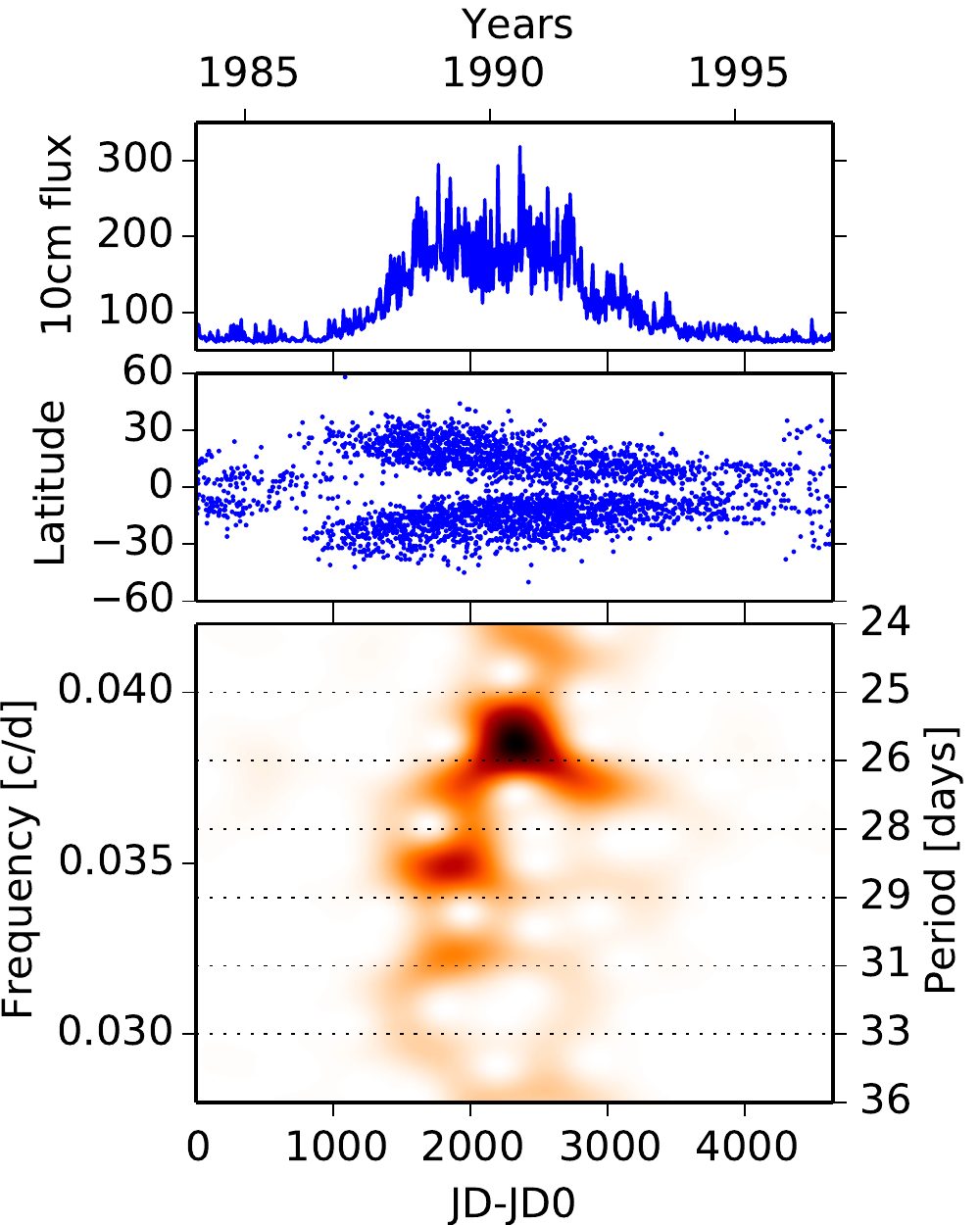}
\caption{Testing the STFT-method on solar radio data. Both plots show the 10\,cm radio flux (top), the butterfly-diagram (time--spot latitude diagram, middle), and the result of the STFT analysis. The low spottedness of the inactive Sun does not makes the possible detection of cycles very uncertain, although some changes in the detected rotation period might seem on the right zoomed-in plot. }
\label{fig:sun-test}
\end{figure*}

\begin{figure*}
\centering
\includegraphics[width=0.45\textwidth]{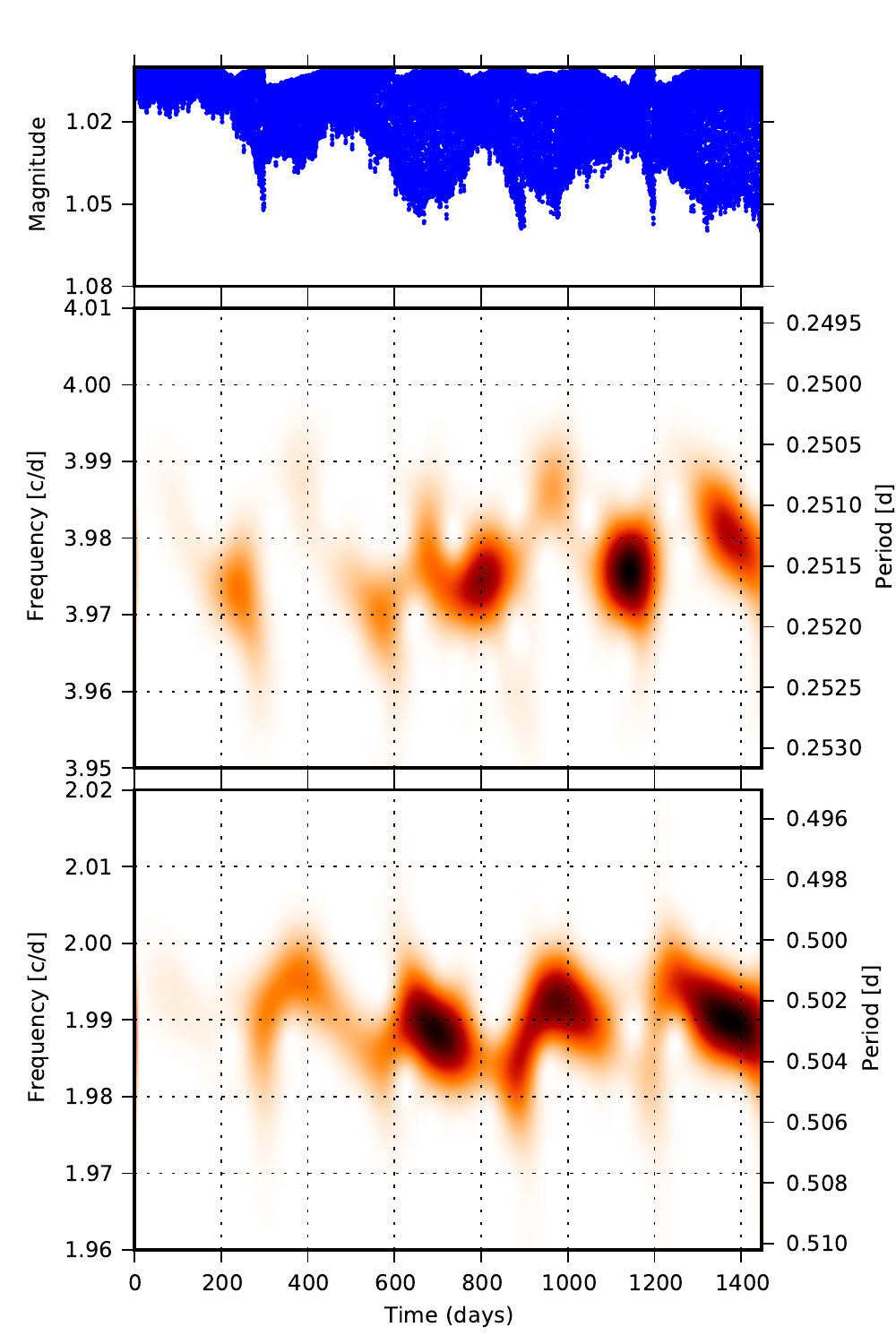}
\includegraphics[width=0.45\textwidth]{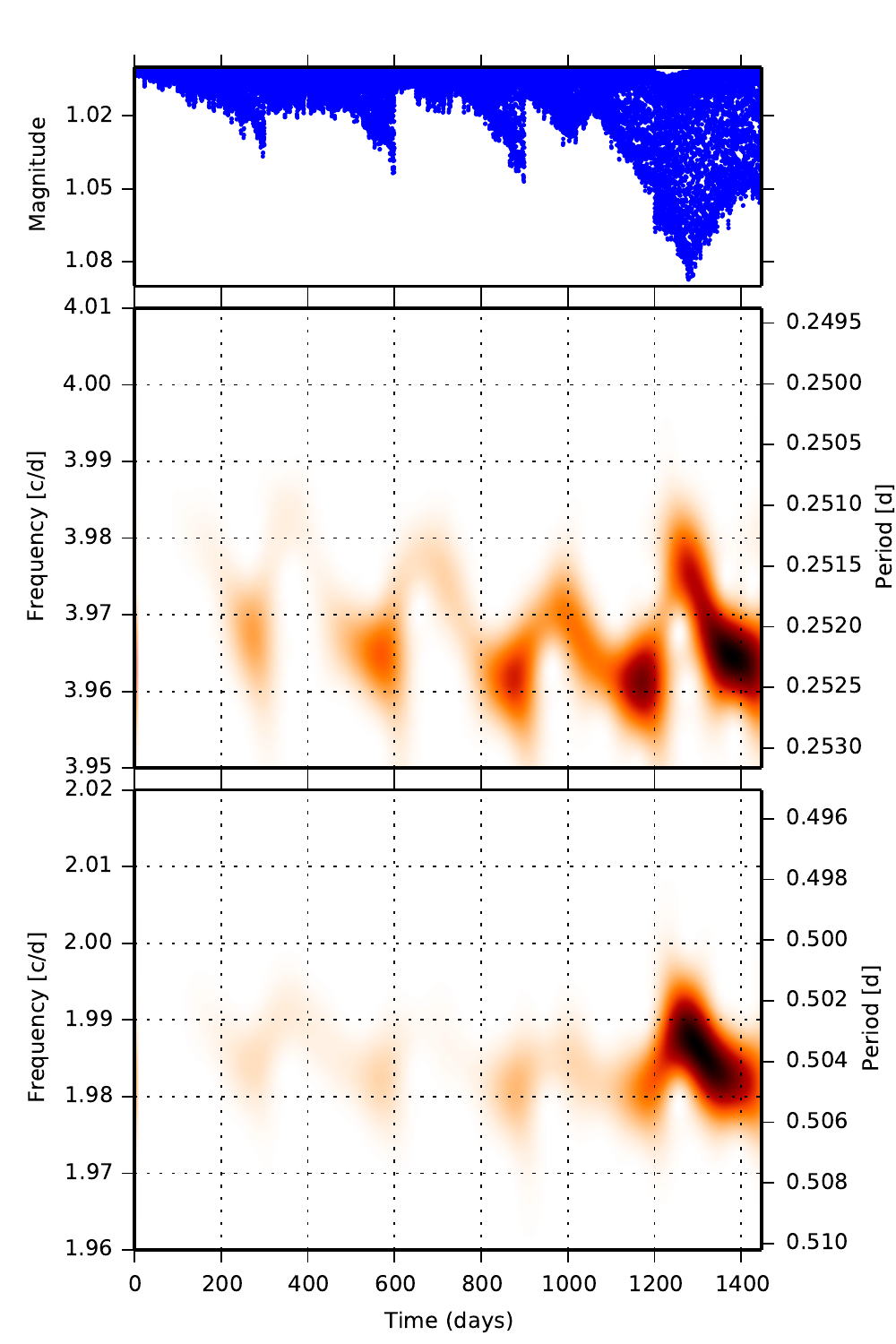}
\caption{Testing the STFT-method on artificial light curves. The cycle length could be recovered in both cases within the errors, although the large amplitude differences in the light curve, and the weighting in the STFT plotted on the right makes the recognizing harder. }
\label{fig:sml-test}
\end{figure*}

\begin{figure*}
\centering
\includegraphics[width=0.45\textwidth]{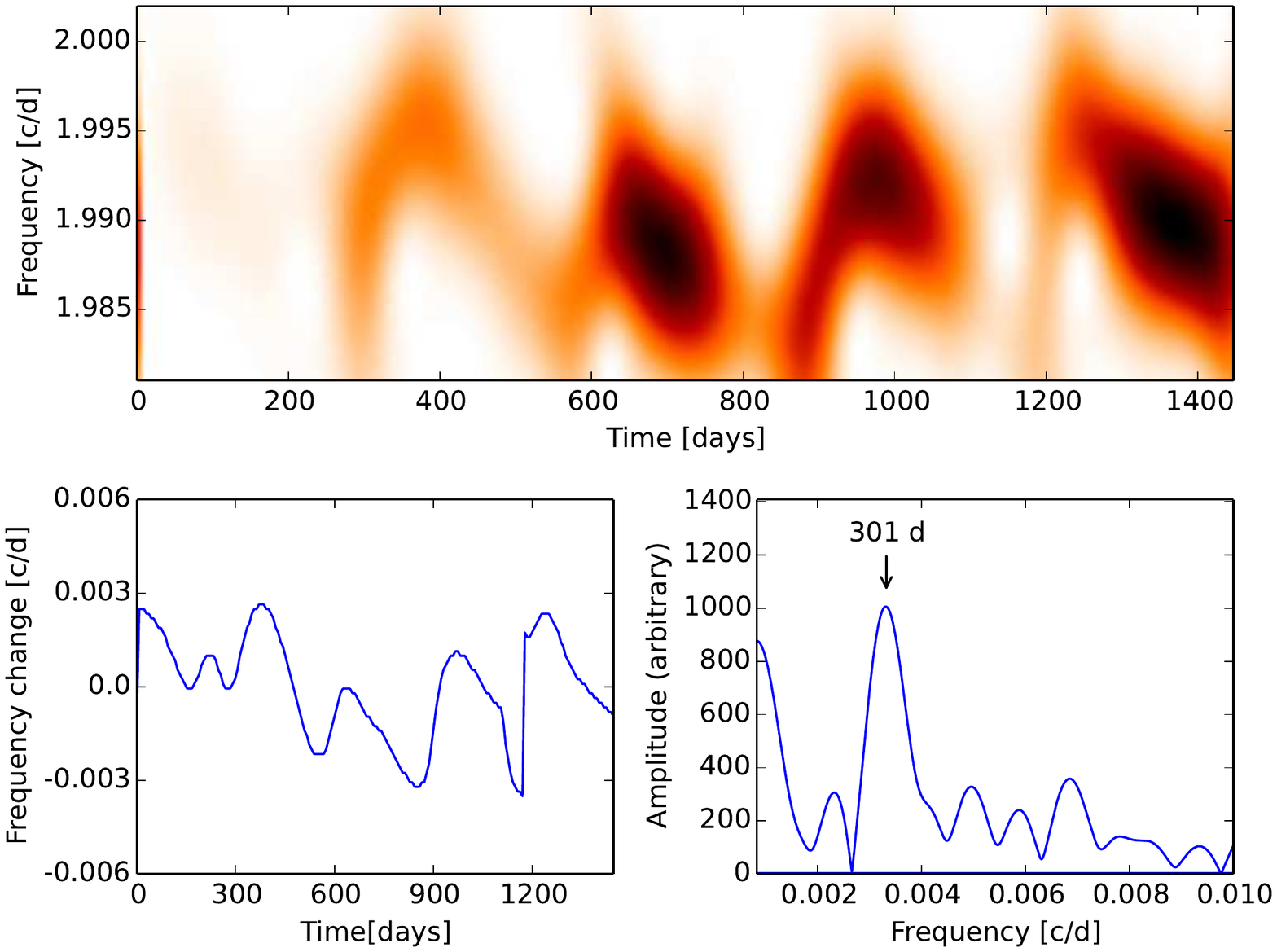}
\includegraphics[width=0.45\textwidth]{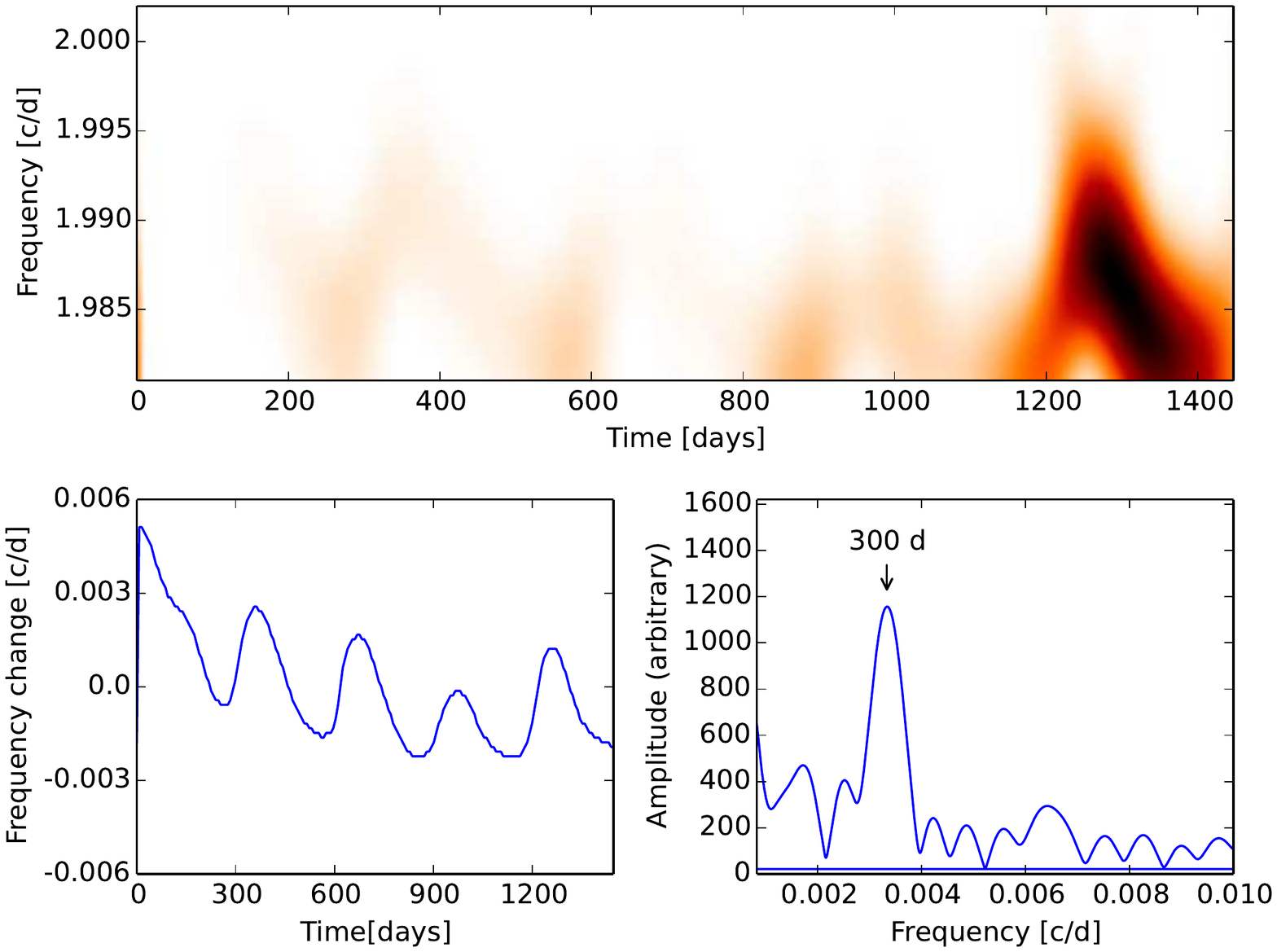}
\caption{Recovering cycle length from artificial data. The light curves are the same as plotted in Fig.~\ref{fig:sml-test}. The results  return correctly the 300 day-long periods.} 
\label{fig:sml-test2}
\end{figure*}

To test our method we applied it to solar observations and artificial light curves. To study the Sun in a similar way to the stars we need disk-integrated observations, e.g. the 10\,cm radio flux. The result of the STFT analysis is shown in Fig.~\ref{fig:sun-test} on the time scale of decades, and also for one activity cycle. Unfortunately for the analysis, the spottedness of the Sun is low. Therefore the changes in the typical rotation frequency are not seen well, although faster rotation is observed when the spots are closer to the equator, in the activity maximum (see the zoomed-in plot of Fig.~\ref{fig:sun-test}), near 28 days.
Note, that by using the 10\,cm solar flux, we are analyzing large coronal structures of the Sun, therefore, the changes in the rotation frequency are not so well pronounced. A very similar time-frequency analysis was performed recently by \cite{sun-sml} using solar Ca {\sc ii} K-data. They used a renormalized plot (see Fig. 10 in the paper) to enhance and show the features that are otherwise lost during solar minimum -- using this method, the quasi-periodic changes through the cycles can indeed be seen.

To evade problems emerging from the low spottedness of the Sun, we tested the method also on artificial data. We used a three-spot model of a fast-rotating star ($P_\mathrm{rot}=0.5d$) with an inclination of $i=50^\circ$ produced by SpotModeL \citep{sml}. During the supposed 300 day-long activity cycle in the data, typical spot latitudes were changing between 30--70$^\circ$ (to simulate the higher spot latitudes due to Coriolis-effect caused by fast rotation). Spot longitudes were chosen randomly, and varied according to solar-like differential rotation with $\alpha=0.01$. When generating the light curves, a small amount of random variation was added to each spot parameter.  Note, that the actual values of the rotation period and cycle length are completely arbitrary in this simulation, they affect only the time scale of the variations, but were chosen to be similar to the values we found in the \textit{Kepler}-stars. We plotted two examples of the results using STFT analysis on these artificial data in Fig.~\ref{fig:sml-test}.
Spot coverage variation was set between 3--10\%. In the first case, our analysis gave a cycle length of  $301\pm40d$, while in the second case the result was $300\pm35d$. In both cases, the supposed 300 day-long cycle length was recovered  correctly.

\bsp

\label{lastpage}


\begin{thebibliography}{99}
\bibitem[Augustson et al.(2013)]{dynamo1} Augustson, K.C., Brun, A.S., Toomre, J., The Astrophysical Journal, 2013,  777, 153 


\bibitem[Baliunas et al.(1996)]{1996ApJ...460..848B} Baliunas, S.L., Nesme-Ribes, E., Sokoloff, D., Soon, W.H., Astrophysical Journal v.460, 1996,  460, 848 

\bibitem[Berdyugina \& Henry(2007)]{2007ApJ...659L.157B} Berdyugina, S.V., Henry, G.W., The Astrophysical Journal, 2007,  659, L157--L160 


\bibitem[Brown et al.(2011)]{magneticwreath} Brown, B.P. et al., The Astrophysical Journal, 2011,  731, 69 

\bibitem[Brun et al.(2004)]{dynamo2} Brun, A.S., Miesch, M.S., Toomre, J., The Astrophysical Journal, 2004,  614, 1073--1098 

\bibitem[Csubry \& Koll\'ath(2004)]{tifran} Csubry, Z.; Koll\'ath, Z., 2006 ASP Conference Series, 349, 215C

\bibitem[Deeg et al.(2004)]{PASS} Deeg, H.~J., Alonso, R., Belmonte, J.~A. et al.\ 2004, PASP, 116, 985 

\bibitem[I{\c s}{\i}k, Schmitt, \& Sch{\"u}ssler(2011)]{emre} I{\c s}{\i}k, E., Schmitt, D., \& Sch{\"u}ssler, M.\ 2011, Astronomy and Astrophysics, 528, A135 

\bibitem[Ivezi\'c et al.(2008)]{lsst}
Ivezi\'c, \v{Z}. et al: \emph{LSST: from Science Drivers to Reference Design and 
Anticipated Data Products}\footnote{Available from http://lsst.org/lsst/overview/}

\bibitem[Karoff et al.(2013)]{kepleract3} Karoff, C. et al., Monthly Notices of the Royal Astronomical Society, 2013,  433, 3227--3238 

\bibitem[Katsova et al.(2010)]{Katsova:2010kt} Katsova, M.M. et al., Proceedings of the International Astronomical Union, 2010,  15, 274--281 

\bibitem[Koll\'ath(1990)]{1990KOTN....1....1K} Koll\'ath, Z., Konkoly Observatory Occasional Technical Notes, 1990,  1, 1 

\bibitem[Koll\'ath \& Ol\'ah(2009)]{2009AA...501..695K} Koll\'ath, Z., Ol\'ah, K., Astronomy and Astrophysics, 2009,  501, 695--702 

\bibitem[Mathur et al.(2013)]{kepleract1} Mathur, S. et al., eprint arXiv:1312.6997, 2013

\bibitem[McQuillan, Aigrain, \& Mazeh(2013)]{keplerrot} McQuillan, A., Aigrain, S., Mazeh, T., Monthly Notices of the Royal Astronomical Society, 2013,  432, 1203--1216 

\bibitem[McQuillan, Mazeh, \& Aigrain(2013)]{keplerrot2} McQuillan, A., Mazeh, T., Aigrain, S., The Astrophysical Journal, 2013,  775, L11 

\bibitem[Nielsen et al.(2013)]{keplerrot3} Nielsen, M.B., Gizon, L., Schunker, H., Karoff, C., Astronomy and Astrophysics, 2013,  557, L10 

\bibitem[Ol\'ah et al.(2009)]{2009AA...501..703O} Ol\'ah, K. et al., Astronomy and Astrophysics, 2009,  501, 703--713 

\bibitem[Ol\'ah, Jurcsik, \& Strassmeier(2003)]{uzlib} Ol\'ah, K., Jurcsik, J., Strassmeier, K.G., Astronomy and Astrophysics, 2003,  410, 685--689 

\bibitem[Ol\'ah, Koll\'ath, \& Strassmeier(2000)]{2000AA...356..643O} Ol\'ah, K., Koll\'ath, Z., Strassmeier, K.G., Astronomy and Astrophysics, 2000,  356, 643--653 

\bibitem[Ol{\'a}h \& Strassmeier(2002)]{rot-cyc} Ol{\'a}h, K., \& Strassmeier, K.~G.\ 2002, Astronomische Nachrichten, 323, 361 

\bibitem[P\'al et al.(2013)]{flyseye}P\'al A. et al., Astronomische Nachrichten, accepted, 2013arXiv1306.6564

\bibitem[Pettersen,Ol\'ah, \& Sandmann(1992)]{Pettersen1992tg} Pettersen, B.R., Ol\'ah, K., Sandmann, W.H., Proceedings of the International Astronomical Union, 1992,  96, 497--504 

\bibitem[Reinhold, Reiners, \& Basri(2013)]{keplerdr} Reinhold, T., Reiners, A., Basri, G., Astronomy and Astrophysics, 2013,  560, A4 

\bibitem[Rib\'arik,Ol\'ah,\& Strassmeier(2003)]{sml} Rib\'arik, G., Ol\'ah, K., Strassmeier, K.G., Astronomische Nachrichten, 2003,  324, 202--214 

\bibitem[Savanov(2012)]{savanov} Savanov, I.S., Astronomy Reports, 2012,  56, 716--721 

\bibitem[Scargle,Keil,\& Worden(2013)]{sun-sml} Scargle, J.D., Keil, S.L., Worden, S.P., The Astrophysical Journal, 2013,  771, 33 

\bibitem[Strassmeier et al.(1997)]{1997PASP..109..697S} Strassmeier, K.~G., 
Boyd, L.~J., Epand, D.~H., \& Granzer, T.\ 1997, Publications of the Astronomical Society of the Pacific, 109, 697 

\bibitem[Strassmeier et al.(2011)]{tifran-cycle} Strassmeier, K.~G., Carroll, T.~A., Weber, M., Granzer, T., Bartus, J., Ol{\'a}h, K., \& Rice, J.~B.\ 2011, Astronomy and Astrophysics, 535, A98 

\bibitem[Vida et al.(2013)]{shortcyc}Vida, K., Kriskovics, L., Ol\'ah, K. 2013, Astronomische Nachrichten, 334, 972

\bibitem[Walkowicz \& Basri(2013)]{kepleract2} Walkowicz, L.M., Basri, G.S., Monthly Notices of the Royal Astronomical Society, 2013,  436, 1883--1895 

\bibitem[Wilson(1968)]{1968ApJ...153..221W} Wilson, O.C., Astrophysical Journal, 1968,  153, 221 

\end{thebibliography}
\end{document}